\documentclass[reprint,amsmath,amssymb,amsfonts,aps,prbl,floatfix,]{revtex4-1}
\usepackage{times}
\usepackage{graphicx}
\usepackage{dcolumn}
\usepackage{bm}
\usepackage{ulem}
\usepackage[colorlinks=true, citecolor=blue, linkcolor=blue, urlcolor=blue]{hyperref}
\usepackage{bbold}
\usepackage{xcolor}
\usepackage{xr}
\definecolor{grayish}{rgb}{0.86,0.86,0.68}
\definecolor{purple}{rgb}{0.6,0,0.6}

\begin{document}

\title{Theory of correlated insulators and superconductivity in twisted bilayer graphene}
\author{Gal Shavit}
\author{Erez Berg}
\author{Ady Stern}
\author{Yuval Oreg}
\affiliation{
 Department of Condensed Matter Physics, Weizmann Institute of Science, Rehovot, Israel 76100
}

\date{\today}

\begin{abstract}
We introduce and analyze a model that sheds light on the interplay between correlated insulating states, superconductivity, and flavor-symmetry breaking in magic angle twisted bilayer graphene. 
Using a variational mean-field theory, we determine the normal-state phase diagram of our model as a function of the band filling. The model features robust insulators at even integer fillings, occasional weaker insulators at odd integer fillings, and a pattern of flavor-symmetry breaking at non-integer fillings.
Adding a phonon-mediated inter-valley retarded attractive interaction, we obtain strong-coupling superconducting domes, whose structure is in qualitative agreement with experiments. 
Our model elucidates how the intricate form of the interactions and the particle-hole asymmetry of the electronic structure determine the phase diagram. It also explains how subtle differences between devices may lead to the different behaviors observed experimentally. A similar model can be applied with minor modifications to other moir\'{e} systems, such as twisted trilayer graphene. 

\end{abstract}

\maketitle

\textit{Introduction.---}
When two graphene layers are stacked at a relative twist angle of $\sim1.1^{\circ}$, the lowest-lying electron bands become exceptionally flat \cite{BistritzermacdonaldPNAS}. 
Recently, this so-called magic angle twisted bilayer graphene (MATBG) emerged as a highly tunable platform to study strongly-correlated physics. 
Correlated insulators (CIs), where interactions induce a gap and suppress transport, were first observed in MATBG at fillings of $\nu=\pm2$ electrons per moir\'{e} unit cell relative to the charge neutrality point (CNP)~\cite{CaoCorrelatedInsulator,CaoUnconventionalSC}.
Later experiments found a CI at $\nu=+3$ \cite{YankowitzTuningMATBG,YoungTuningSC}, and in some instances CIs were measured at nearly all integer fillings \cite{EfetovAllIntegers}.
Empirically, insulating behavior is more pronounced for electrons ($\nu>0$).
The origin of these integer-filling CIs has been explored in several recent works~\cite{SenthilAshvin2018PRX,Dodaro2018,Kang2019,UchoaQuarterfilling,BultnickKhalaf2020,Kang2020,VAFEKRG,TBG4Exactinsulatorgroundphasediagram,TBG6exactdiagonalizationintegerfillings}.

Another remarkable feature of MATBG is the appearance of superconducting domes near the CIs at $\nu=\pm 2$~\cite{CaoUnconventionalSC,YankowitzTuningMATBG,YoungTuningSC}, with superconductivity generally being more robust for holes ($\nu<0$), and (for both electrons and holes) on the $\left|\nu\right|>2$ side. 
Experiments manipulating the electrostatic screening have indicated that Coulomb repulsion is either detrimental to superconductivity in MATBG or weakly affects it~\cite{YankowitzTuningMATBG,EfetovTuningSC,YoungTuningSC,BLGscreening}. 
This suggests that electron-phonon coupling may play a role in MATBG \cite{MacdonaldPRLopticalPhonons,LianBioBernevigPRLphonons,Wu2019,TBG5phononsNotCoulomb,Ruhman2021}, and plausibly induce superconductivity at certain fillings.
However, the interplay between strong repulsion and its effect on the normal-state, retarded attraction due to phonons, and the unique multi-band structure have yet to be fully explored.

In this manuscript, we introduce and investigate a phenomenological model and find that it exhibits the most salient features of MATBG observed in experiments.
The model comprises four electronic ``flavors'', accounting for spin and valley
degeneracies, and interactions with strengths of the order of their bandwidth.
The structure of the interaction terms and the features of the density of states (DOS) of non-interacting MATBG determine the phase diagram.

\begin{figure*}
\begin{centering}
\includegraphics[scale=0.42]{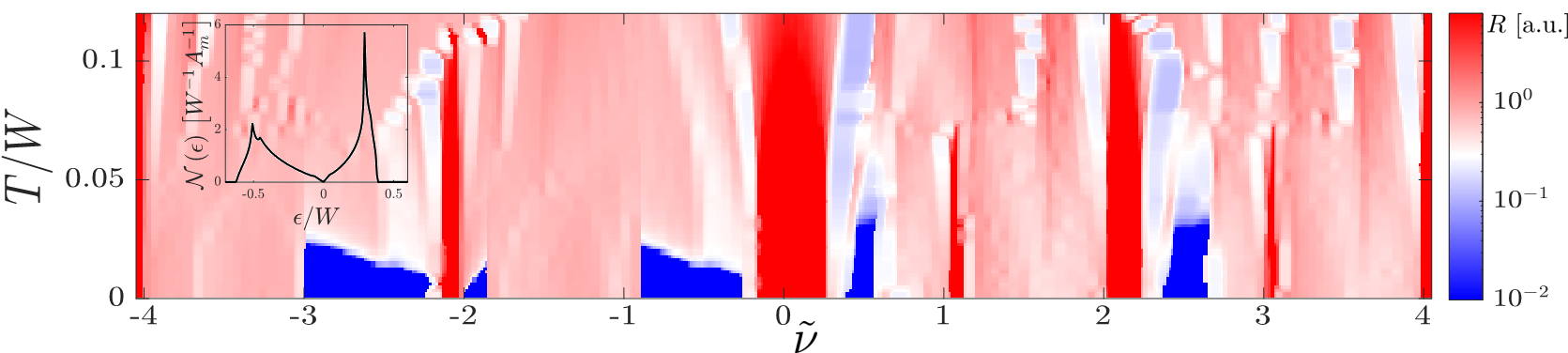}
\par\end{centering}
\caption{\label{fig:schematics} Temperature and filling ($\tilde \nu$, see SM Sec.~S.3) phase diagram of the model. We plot the resistivity, inversely proportional to the compressibility far from the superconducting phase, see SM Sec.~S.5. The model features CIs near certain integer fillings and superconductivity, both in qualitative agreement with experiments. We used the interaction parameters [Eq.~\eqref{eq:Hint}] $U_{{\rm C}}=0.7W$, $U_{\delta}=0.15W$, $g_1=g_2=0.12W$, and phonon-mediated attraction strength [Eq.~\eqref{eq:RGequation}] $V^{*}=0.24W$. Inset: DOS of the single-particle Hamiltonian~\eqref{eq:singleParticle}.
For a detailed schematic phase diagram,
and the effect of weaker Coulomb interactions, see SM.}
\end{figure*}

We find electron correlations induce CIs at even-integer fillings with inter-valley coherent (IVC) order (i.e., spontaneously breaking valley $U_{\rm v}\left(1\right)$ symmetry), whereas the odd-integer CIs, typically having bands with non-zero Chern numbers, are more sensitive to details of sub-leading interaction terms.
At non-integer fillings, the system is not fully gapped, yet, the spin-valley flavor symmetry is broken ~\cite{YazdaniRevivals,DiracRevivals,kang2021cascadesVafekBernevig}. 
Retarded inter-valley attractive interactions, due to e.g., phonons~\cite{LianBioBernevigPRLphonons}, then enable the formation of superconducting domes, which are most prominent at fillings which agree remarkably well with experiments.
As depicted in Fig.~\ref{fig:schematics}, we recover a superconducting dome flanked by two insulators near $\nu=+2,+3$, and a more substantial dome on the hole-doped side of the $\nu=-2$ CI.

At certain fillings, strong-coupling superconductivity may be established, i.e., $T_c$ becomes an appreciable fraction of the Fermi temperature $T_F$, leading to significant superconducting phase fluctuations, whose effect on transport we account for.
This is enabled by the underlying normal state, where interactions induce spontaneous breaking of flavor-symmetry breaking \textit{and} the valley $U_{\rm v}\left(1\right)$ symmetries.
Moreover, this symmetry-broken state has only two active flavors in different valleys and opposite spins, hence it may sustain large in-plane magnetic fields.


\textit{Model and results.---}
Our model comprises eight flat with valley ($K/K'$), spin, and sublattice ($A/B$) degrees of freedom, labeled by Pauli matrices $\tau_{i}$, $s_{i}$,  and $\sigma_{i}$, respectively.
This basis is motivated by the MATBG sublattice-polarized basis discussed in Ref.~\cite{BultnickKhalaf2020}. 
These bands have a valley-dependent Chern number, $C=\tau_{z}\sigma_{z}$.
The model Hamiltonian is
\begin{equation}
H=\sum_{\mathbf{k}}\Psi_{\mathbf{k}}^{\dagger}h_{0}\left(\mathbf{k}\right)\Psi_{\mathbf{k}}+H_{{\rm int}},\label{eq:ModelHamiltonian}
\end{equation}
where $H_{{\rm int}}$ describes the interactions,
$\Psi_{\mathbf{k}}$ is an 8-spinor of fermionic operators $c_{\tau s\sigma}\left(\mathbf{k}\right)$ (annihilating an electron at valley $\tau$, spin $s$ and sub-lattice $\sigma$), and 
\begin{equation}
h_{0}\left(\mathbf{k}\right)=f_{x}\left(\mathbf{k}\right)\sigma_{x}+f_{y}\left(\mathbf{k}\right)\sigma_{y}\tau_{z} + f_{\rm p-h}\left(\mathbf{k}\right).\label{eq:singleParticle}
\end{equation}
The functions $f_x$, $f_y$, and $f_{\rm p-h}$ determine the dispersion in the moir\'{e} Brillouin zone (mBZ), which has two Dirac cones with the same chirality, and reproduces an electronic DOS with the prominent features of the MATBG flat-bands (see Fig.~\ref{fig:schematics} inset).
Namely, linearly increasing DOS near the CNP, pronounced DOS peak near half-filling of the conduction/valence bands followed by a decline towards the band edge, and appreciable particle-hole asymmetry.
The combined bandwidth of the conduction and valence bands in the mBZ is $W$.
We note the form of $h_{0}\left(\mathbf{k}\right)$ preserves $C_2=\tau_x\sigma_x$ and time-reversal symmetries ${\cal T}=\tau_x{\cal K}$, with ${\cal K}$ the complex-conjugation operator~\cite{C3h0}.
For more about $h_{0}\left(\mathbf{k}\right)$, see Supplementary Materials (SM), Sec.~S.1~\cite{SupplementRef}.

We write electron-electron interactions as a sum of local interaction terms,
\begin{equation}
H_{{\rm int}}=\sum_{\alpha,\mathbf{k},\mathbf{k'},\mathbf{q}}\frac{\lambda_{\alpha}}{2\Omega}\left(\Psi_{\mathbf{k+q}}^{\dagger}\vec{{\cal O}}_{\alpha}\Psi_{\mathbf{k}}\right)\cdot\left(\Psi_{\mathbf{k'-q}}^{\dagger}\vec{\cal O}_{\alpha}\Psi_{\mathbf{k'}}\right),\label{eq:Hint}
\end{equation}
where $\Omega$ is the volume, $\vec{{\cal O}}_{\alpha}$ are matrices in valley-spin-sublattice space, and~$\lambda_{\alpha}$ are coupling constants.
The dominant term is the density-density interaction with $\vec{\cal O}_{1}=\mathbb{1}$, $\lambda_{1}=U_{{\rm C}}$, and reflects the screened Coulomb repulsion. 
We consider a secondary interaction $\vec{\cal O}_{2}=\left(\tau_z \sigma_x,\sigma_y\right)$ with $\lambda_{2}=U_{\delta}$, accounting for the form-factors obtained when projecting the Coulomb repulsion onto the flat-bands away from the chiral limit~\cite{BultnickKhalaf2020}.
Additional terms are inspired by instantaneous interactions due to electron-optical-phonon interactions, 
$\vec{\cal O}_{3}=\left(\sigma_y\tau_z,\sigma_x\right)$ with $\lambda_{3}=g_{1}$, and  $\vec{\cal O}_{4}=\left(\tau_x\sigma_x,\tau_y\sigma_x\right)$ with $\lambda_{4}=g_{2}$.
Their structure is dictated by the electron-phonon coupling to low-momentum phonons ($\vec{\cal O}_{3}$) and to valley-momentum phonons ($\vec{\cal O}_{4}$)~\cite{MacdonaldPRLopticalPhonons}.
The phonon-induced interactions are attractive, i.e., $g_1,g_2<0$.
The interactions preserve $C_2$, $\cal{T}$, and $C_{3}=e^{2\pi i/3\sigma_{z}\tau_{z}}$ symmetries
~\cite{check}. 

We study the model~\eqref{eq:ModelHamiltonian}--\eqref{eq:Hint} using a variational Hartree-Fock approach. We minimize the grand-potential $\Phi$,
at a given temperature $T$ and chemical potential $\mu$, generated by the variational Hamiltonian $H_{{\rm MF}}=\sum_{\mathbf{k}}\Psi_{\mathbf{k}}^{\dagger}h_{{\rm MF}}\left(\mathbf{k}\right)\Psi_{\mathbf{k}}$~\cite{SupplementRef}.
We note that in the mean-field approach, due to the local nature of the interactions, the details of the non-interacting dispersion do not play a role, only the DOS. 
We explore three kinds of spontaneous symmetry breaking in $h_{{\rm MF}}$:
(i) Flavor-symmetry breaking, i.e., one or several of the operators $\left\{s_{z},\tau_{z},\tau_{z}s_{z}\right\} $ attain a finite expectation value; 
(ii) intra-flavor sublattice-symmetry breaking ($\sigma_{z}$ terms), leading to Chern gaps; 
(iii) IVC order with a finite expectation value for $\tau_{x}\cos\gamma+\tau_{y}\sin\gamma$.
We restrict our analysis to IVC terms of the form
\begin{equation}\label{eq:ivcorder}
\Delta_{{\rm ivc}}^{\pm}\frac{1\pm s_{z}\tau_{z}}{2}\tau_{y}s_{x}\sigma_{y}.
\end{equation}
This order-parameter resembles the Kramers-IVC of Ref.~\cite{BultnickKhalaf2020}, with an effective time-reversal symmetry ${\cal T}'=\tau_y s_x {\cal K}$. 
The choice of $s_{x}\sigma_{y}$ in \eqref{eq:ivcorder} is justified \textit{a posteriori} by examination of the mean-field interaction energy (see SM Sec.~S.1~\cite{SupplementRef}).
We find that the $g_2$ interaction favors orders where the spin is anti-aligned in opposite valleys, justifying~$s_{x}$ in Eq.~\eqref{eq:ivcorder}.
Moreover, we find $U_{\delta}$ and $g_1$ favor states where 
$\left\langle c_{\tau s\sigma}^{\dagger}c_{\bar{\tau}\bar{s}\bar{\sigma}}\right\rangle =-\left\langle c_{\tau s\bar{\sigma}}^{\dagger}c_{\bar{\tau}\bar{s}\sigma}\right\rangle ^*$,
so IVC orders $\propto\sigma_y$ gain interaction energy.
Lastly, our analysis suggests sublattice-symmetry breaking is favored by $g_1$, yet \textit{suppressed} by $U_{\delta}$. 
The interplay between these interactions is key to understanding why insulators at odd fillings are experimentally less robust than those at even fillings.

\begin{figure}
\begin{centering}
\includegraphics[scale=0.53]{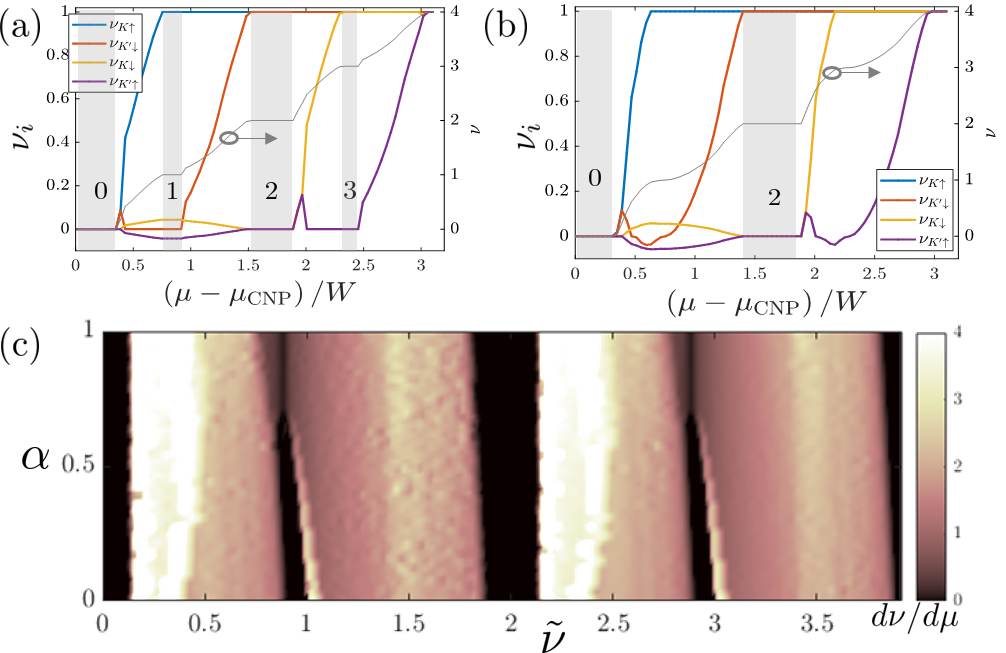}
\par\end{centering}
\caption{\label{fig:normalPhaseDiagram} (a) $T=0$ mean-field occupation $\nu_{i}$ of individual flavors and total filling $\nu$ as a function of chemical potential.
Grey rectangles mark incompressible regions.
Interaction parameters used: $U_{{\rm C}}=0.75W$, $U_{\delta}=g_1=0.1W$, $g_2=0.08W$. 
(b) Same as (a), with $U_{\delta}=0.2W$ and $g_1=0$.
(c) Compressibility $d\nu/d\mu$ as a function of $\tilde\nu$ and $\alpha\equiv\frac{U_{\delta}-g_1}{U_{\delta}+g_1}$,
retaining $U_{\delta}+g_1=0.2W$, and all other parameters from (a)--(b). 
Notice (a) corresponds to $\alpha=0$ and (b) to $\alpha=1$.}
\end{figure}

Mean-field phase diagram results are displayed in Fig.~\ref{fig:normalPhaseDiagram}.
Panels (a)--(b) show the filling $\nu_{i}$ of each flavor for different values of $U_{\delta}$ and $g_1$. 
Our results feature a sequence of symmetry-breaking phase transitions. At the CNP, the system is in a fully-gapped IVC state.
With increased~$\mu$, the IVC gap in one $\tau_z s_z$ sector closes, and the two flavors making up that sector begin to populate [near $\left(\mu-\mu_{\rm CNP}\right)/W\approx 0.4$ in Fig.~\ref{fig:normalPhaseDiagram}].
This is followed by flavor-symmetry breaking within that sector,
where one flavor is depleted and the other is filled.
Depending on details, the depleted flavor may develop a Chern gap,
leading to an incompressible region near $\nu=1$. 
Increasing $\mu$ further, this flavor is gradually filled.
This process repeats for the flavors in the other IVC sector (starting at $\nu=2$), following an incompressible regime, where two flavors are full, and two others are IVC-gapped.

We note that in a region around $\nu=1$, flavor-polarization develops in the IVC sector, 
yet it remains incompressible.
This is due to spin-polarization in the more populated sector,
promoting opposite polarization in opposing valleys, gaining interaction energy $\propto\left|g_2\right|$.

In Fig.~\ref{fig:normalPhaseDiagram}(c) we plot the compressibility $d\nu/d\mu$ as a function of $\alpha\equiv\frac{U_{\delta}-g_1}{U_{\delta}+g_1}$ and $\tilde {\nu}$. 
The latter is a proxy for the filling fraction representing the experimental scenario, where a back-gate voltage tunes the electron filling, see SM Sec.~S.3~\cite{SupplementRef}.
As $\alpha$ increases, ($g_1$ becomes smaller compared to $U_{\delta}$) the odd-filling gaps close and eventually vanish at $\alpha\sim0.7$, giving way to finite but low compressibility~\cite{DiracRevivals}. 
This trend agrees with our analytical examination of the roles of $U_\delta$ and $g_1$.
The $\tilde \nu=0,2$ incompressible IVC states weakly depend on $\alpha$, and thus expected to be more robust.

The phase diagram establishes that the appearance of CIs either at all integer fillings, or only at even ones, depends delicately on the details and hierarchy of the effective interaction terms~\cite{QHFM}. 
We note that the appearance of $\sigma_y$--IVC orders at even fillings agrees with the predictions of Ref.~\cite{BultnickKhalaf2020} and was verified numerically~\cite{hofmann2021fermionic}.
This is expected as the $U_{\delta}$ term captures the effect of the density form-factors of the projected interaction. 
Our model thus provides a tractable way of going beyond specific integer fillings and tracking the evolution of the mean-field ground-state with $\mu$.

\textit {Superconductivity.---} 
Our starting point of examining superconductivity in MATBG is the symmetry-breaking cascade obtained above. 
We explore inter-valley pairing mediated by electron-phonon interactions. 
The inter-valley pairing is favored both by the acoustic phonons~\cite{LianBioBernevigPRLphonons} and since intra-valley Cooper pairs have finite-momentum.
Thus, we focus on valley-degenerate areas in the phase diagram.
We note that scenarios where the superconducting condensation energy gain is sufficient to depolarize opposite-valley flavors are not considered.

As discussed, the model favors inter-valley antiferromagnetism, naturally preferring opposite-spin pairing.
Restricting our discussion to the simplest scenario where the pairing lacks sublattice-structure (it is sufficient to capture the most salient experimental features), we study the pairing amplitude $\Delta_{\tau s}=\Delta_{\tau s A}=\Delta_{\tau s B}$, where
\begin{equation}
\Delta_{\tau s \sigma}=\frac{1}{\Omega}\sum_{\mathbf{k}}\left\langle c_{\bar{\tau}\bar{s}\sigma}\left(-\mathbf{k}\right)c_{\tau s\sigma}\left(\mathbf{k}\right)\right\rangle .\label{eq:PairingAmplitude}
\end{equation}
We note that due to the aforementioned spontaneous spin-valley locking and flavor-symmetry breaking, the system attains non-zero spin-triplet pairing correlations~\cite{KTLawPairingCorrelationsTriplet}. 
This may lead to phenomenology similar to that of Ising superconductors, namely a critical in-plane magnetic field that is set by the normal-state energetics, exceeding the Pauli-Chandrasekhar-Clogston limit~\cite{ClogsonLimit,IsingSuperconductivityMoS2,IsingSuperconductivityNbSe2}.

Adopting a Tolmachev-Morel-Anderson renormalization group (RG) approach~\cite{tolmachev1962logarithmic,AndersonMorel}, we account for the effects of Coulomb repulsion as well as the phonon-mediated attraction. 
Neglecting the attraction for now, we begin with the action ${\cal S}={\cal S}_{{\rm MF}}+{\cal S}_{{\rm Cooper}}$, where $S_{{\rm MF}}$ is corresponds to the solution of the variational procedure, and 
${\cal S}_{{\rm Cooper}}=\int d^{2}\mathbf{x}\sum_{\tau s\sigma}c_{\tau s\sigma}^{\dagger}c_{\bar{\tau}\bar{s}\sigma}^{\dagger}\left[\frac{U_{{\rm C}}}{2}c_{\bar{\tau}\bar{s}\sigma}c_{\tau s\sigma}-\left(U_{\delta}+\left|g_{1}\right|\right)c_{\bar{\tau}\bar{s}\bar{\sigma}}c_{\tau s\bar{\sigma}}\right]$
is the interaction in the $\Delta_{\tau s}$ Cooper channel. 
Following the standard RG procedure~\cite{nagaosa1999quantum,SupplementRef}, we find the flow of the coupling constant $V$ as a function of the energy cutoff $\Lambda$. 
The initial conditions are $\Lambda_{0}=W$, and $V_{0}=\frac{U_{{\rm C}}}{2}-\left(U_{{\rm \delta}}+\left|g_{1}\right|\right)$.
Notice the secondary interactions \textit{enhance} pairing
whereas Coulomb repulsion suppresses it.

We now address the role of the acoustic phonon branch mediating the retarded attraction.
We observe that due to folding of the phonon spectrum into the mBZ~\cite{GuineaFolding,FoldedPhonons}, one should also consider generated ``pseudo-optical'' branches.
Consequently, the RG equation for the coupling constant is \cite{SupplementRef} \begin{equation}
\frac{d}{d\Lambda}V=\frac{{\cal N}\left(\Lambda\right)}{\Lambda}V^{2}+\frac{V^{*}}{W},\label{eq:RGequation}
\end{equation}
where the conventional RG flow yields the first term, with ${\cal N}\left(\Lambda\right)$ the electronic DOS.
The non-standard second term appears because as one lowers the cutoff, more phonon modes become attractive, we denote their total contribution by $V^*$, see SM, Sec.~S.4.

\begin{figure}
\begin{centering}
\includegraphics[scale=0.55]{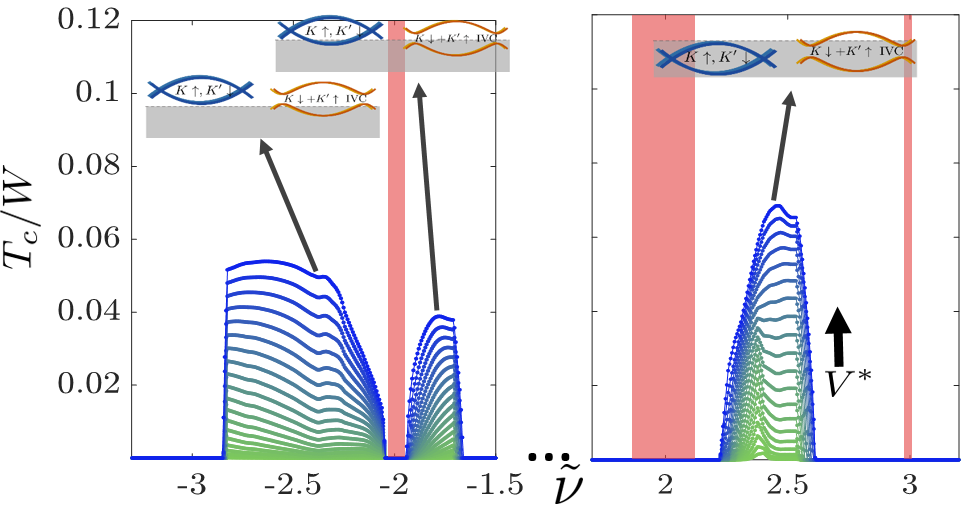}
\par\end{centering}
\caption{\label{fig:Superconductivitymap}
 Superconducting $T_c$ enhancement with increasing retarded attraction $V^*$ near representative fillings. We used the  parameters of Fig.~\ref{fig:schematics}, with $V^* \in \left[0.12,0.32\right]W$, in steps of $0.01$W. Red: incompressibility regions. Direction of increased $V^*$ is indicated, as are the schematic mean-field states from which superconductivity emerges.
}
\end{figure}

Using Eq.~\eqref{eq:RGequation}, in conjunction with the mean-field results, we find $T_{c}$, extracted as the scale at which the coupling constant becomes comparable with the bandwidth, $\left|V\left(T_{c}\right)\right|=W$, signaling its divergence.
Notice that because $W$ is the scale at which retarded phonons begin to contribute, at a given $V^*$ and $V_0$, Eq.~\eqref{eq:RGequation} may lead to a critical $W$, \textit{below which}  superconductivity is lost. This is due to the retardation being ineffective in changing the sign of $V$ along the shorter RG flow.
Fig.~\ref{fig:Superconductivitymap} tracks the evolution of superconductivity domes with increasing phonon-mediated attraction $V^*$.

To take into account the effects of superconducting phase fluctuations on transport, which may be significant
as $T_c$ and $T_F$ become comparable, we use the Halperin-Nelson formula to calculate the resistivity, see SM Sec.~S.5~\cite{SupplementRef,HalperinSuperconductorResistance}.
The difference between the mean-field $T_c$ and the BKT temperature can be parameterized by $T_{\rm BKT}=T_c/\left(1+\tau_c\right)$, where $\tau_c\approx T_c/T_F$.
Experimental estimates~\cite{CaoUnconventionalSC} of $\tau_c\sim0.05-0.1$ in MATBG are in qualitative agreement with the values obtained for $T_c$ and $T_F$ with our model, where we find $\tau_c$ in a similar range on the hole side of Fig.~\ref{fig:schematics}, and $\tau_c$ reaching up to $\sim0.3$ on the electron side.
Non-zero normal-state $\Delta_{\rm ivc}^\pm$ modifies the dispersion, enabling an appreciable Fermi-level DOS even at minuscule fillings. It thus contributes to increasing $T_c/T_F$ as compared to what is expected from a Dirac-like dispersion.

Fig.~\ref{fig:schematics} features three superconducting domes on the hole side. 
The most prominent one occurs at $\tilde {\nu}=-2-\epsilon$, bordering the $\tilde {\nu}=-2$ IVC phase.
Here, the gap in the IVC sector gradually closes with decreasing $\tilde{\nu}$, until it vanishes. 
The suppression in superconductivity near $\tilde{\nu}\approx -3$ occurs due to flavor-polarization, similar to those shown in Fig.~\ref{fig:normalPhaseDiagram}.
Superconductivity at this filling is the most experimentally robust,
often observed with similar double-hump shape~\cite{CaoUnconventionalSC}.
This shape is due to the two-step process, where first the IVC gap closes with doping, and then two gapless flavors get populated.
A similar, but narrower and higher dome emerges at $\tilde \nu=2+\epsilon$. This is because the electron side has larger DOS leading to stronger effective repulsion and wider regions with flavor-polarization. On the other hand, without polarization the larger DOS leads to higher $T_c$. 

A secondary superconducting feature observed in some experiments appears near $\left|\nu\right|=2-\epsilon$, and is also manifest in our model at $\tilde \nu=-2+\epsilon$. 
Its existence is due to depolarization of the two non-IVC flavors when $\tilde{\nu}$ is decreased (see SM, Fig.~S1), resulting from the drop in DOS near the band edge. Thus, this feature is sensitive to the flat-band dispersion details, possibly explaining its haphazard occurrence.

Lastly, we find superconductivity near the CNP, seldom observed in experiments~\cite{EfetovAllIntegers}. 
Essentially, it is a modified copy of the $\tilde {\nu}=-2-\epsilon$ dome, with two flavors facilitating the pairing, and two forming a gapped-IVC state. It has an electron-side counterpart, too.

\textit {Discussion.---}
We presented a simple phenomenological model unifying key features of MATBG, namely interaction-induced CIs at integer fillings, flavor-symmetry-breaking phase transitions, and non-standard superconductivity, and demonstrating their interplay.
Though we neglect ingredients known to be found in MATBG, i.e., long-range Coulomb interactions, intricate wave-functions, fragile topology, and filling-dependent band-structure, much of the phenomenology is remarkably reproduced.
Our minimalistic description of the system's degrees-of-freedom, and the hierarchy of the interaction energy scales, help to comprehend the experimental phase diagram and its variations between different devices.

The model incorporates two important effects of the twist-induced moir\'{e} lattice.
First, generation of a flat-band dispersion, with greatly enhanced DOS~\cite{BistritzermacdonaldPNAS}. 
Second, a dramatic increase of the electron-phonon coupling~\cite{LianBioBernevigPRLphonons}. 
The large DOS enhances the effects of both electron-electron repulsion and the effective phonon-mediated attraction.
Here, we use a mean-field approach combined with RG method to study the interplay of the two.
Within this paradigm, one expects that the CIs and superconductivity compete with one another.
This is consistent with experiments where the strength of the Coulomb interaction is tuned by manipulating the screening environment~\cite{YoungTuningSC,EfetovTuningSC,BLGscreening}.
Another side-effect of this interplay 
is spontaneous spin-valley locking, e.g., near $\nu=-2-\epsilon$, that may be weakly sensitive to parallel in-plane magnetic fields, leading to a superconducting order parameter with appreciable spin-triplet pairing correlations.

It is worth noting the discrepancies between our simplified model and experimental observations. 
In most experiments, the CNP phase appears semi-metallic (though there are notable exceptions). 
Here, we find the strongest CI at this filling.
Furthermore, we find that a $\left|\nu\right|=3$ insulator is accompanied by an insulator at $\left|\nu\right|=1$,  seldom seen in experiments.
One possible cause is a modification of the band-structure itself the filling changes. 
It has been argued~\cite{BandStructurenoramlizations,HartreeBandStructure,Mcdonaldweakfieldhall,FillingDependentRenormalization} that the flat-bands are least flat near charge-neutrality, which may explain the empirical scarcity of insulators at low fillings.
Another possibility is that the semi-metal at the CNP is promoted by strain~\cite{parker2020straininducedZalatel}.
These effects are not considered in this work.
Moreover, there is convincing experimental evidence~\cite{ISospinPomeranchuk,IlaniPomeranchuk} that flavor-fluctuations near $\left|\nu\right|=1$ are non-negligible, suggesting one should include spin and valley fluctuations to fully understand this regime.

Disorder was also not explored in this model.
As was suggested~\cite{AliceaDisorder}, it may settle the discrepancy regarding the CNP insulator, as well as the absence of a quantized transverse response at odd fillings CIs.
Our proposed framework can help elucidate the roles of both fluctuations (treating our phase diagram as a saddle-point around which fluctuations occur) and disorder (quantifying the competition between phases and accounting for how disorder affects it).  

Our model may be used to investigate additional superconducting channels, e.g., d-wave~\cite{MacdonaldPRLopticalPhonons}, and explore under what conditions they become dominant.
Furthermore, this scheme, with different interactions, single-particle terms, or DOS, 
may apply to other moir\'{e} platforms displaying correlation-induced phenomena, e.g.,
ABC-stacked trilayer graphene on hexagonal boron-nitride (hBN)~\cite{ABChbn}, twisted double-bilayer graphene~ \cite{TwistedDoubleBilayerYankowitz,TwistedDoubleBilayerShen2020}, MATBG aligned with hBN~\cite{hBNgoldhaberGordon,hBNyoung} (where we may explain the absence of superconductivity, SM Sec.~S.6), and  magic-angle twisted trilayer graphene~\cite{TrilayerPablo,TrilayerKim}.

\begin{acknowledgments}

We acknowledge enlightening discussions with Pablo Jarillo-Herrero, Shahal Ilani, Uri Zondiner, Ohad Antebi, and Keshav Pareek.
This project was partially supported by grants from the ERC under the European Union’s Horizon 2020 research and innovation programme (grant agreements LEGOTOP No. 788715 and HQMAT No. 817799), the DFG (CRC/Transregio 183, EI 519/7-1), the BSF and NSF (2018643), the ISF Quantum Science and Technology (2074/19), and a research grant from Irving and Cherna Moskowitz.
\end{acknowledgments}

\bibliography{MATBG}

\begin{thebibliography}{65}%
\makeatletter
\providecommand \@ifxundefined [1]{%
 \@ifx{#1\undefined}
}%
\providecommand \@ifnum [1]{%
 \ifnum #1\expandafter \@firstoftwo
 \else \expandafter \@secondoftwo
 \fi
}%
\providecommand \@ifx [1]{%
 \ifx #1\expandafter \@firstoftwo
 \else \expandafter \@secondoftwo
 \fi
}%
\providecommand \natexlab [1]{#1}%
\providecommand \enquote  [1]{``#1''}%
\providecommand \bibnamefont  [1]{#1}%
\providecommand \bibfnamefont [1]{#1}%
\providecommand \citenamefont [1]{#1}%
\providecommand \href@noop [0]{\@secondoftwo}%
\providecommand \href [0]{\begingroup \@sanitize@url \@href}%
\providecommand \@href[1]{\@@startlink{#1}\@@href}%
\providecommand \@@href[1]{\endgroup#1\@@endlink}%
\providecommand \@sanitize@url [0]{\catcode `\\12\catcode `\$12\catcode
  `\&12\catcode `\#12\catcode `\^12\catcode `\_12\catcode `\%12\relax}%
\providecommand \@@startlink[1]{}%
\providecommand \@@endlink[0]{}%
\providecommand \url  [0]{\begingroup\@sanitize@url \@url }%
\providecommand \@url [1]{\endgroup\@href {#1}{\urlprefix }}%
\providecommand \urlprefix  [0]{URL }%
\providecommand \Eprint [0]{\href }%
\providecommand \doibase [0]{http://dx.doi.org/}%
\providecommand \selectlanguage [0]{\@gobble}%
\providecommand \bibinfo  [0]{\@secondoftwo}%
\providecommand \bibfield  [0]{\@secondoftwo}%
\providecommand \translation [1]{[#1]}%
\providecommand \BibitemOpen [0]{}%
\providecommand \bibitemStop [0]{}%
\providecommand \bibitemNoStop [0]{.\EOS\space}%
\providecommand \EOS [0]{\spacefactor3000\relax}%
\providecommand \BibitemShut  [1]{\csname bibitem#1\endcsname}%
\let\auto@bib@innerbib\@empty
\bibitem [{\citenamefont {Bistritzer}\ and\ \citenamefont
  {MacDonald}(2011)}]{BistritzermacdonaldPNAS}%
  \BibitemOpen
  \bibfield  {author} {\bibinfo {author} {\bibfnamefont {R.}~\bibnamefont
  {Bistritzer}}\ and\ \bibinfo {author} {\bibfnamefont {A.~H.}\ \bibnamefont
  {MacDonald}},\ }\href {\doibase 10.1073/pnas.1108174108} {\bibfield
  {journal} {\bibinfo  {journal} {Proceedings of the National Academy of
  Sciences}\ }\textbf {\bibinfo {volume} {108}},\ \bibinfo {pages} {12233}
  (\bibinfo {year} {2011})},\ \Eprint
  {http://arxiv.org/abs/https://www.pnas.org/content/108/30/12233.full.pdf}
  {https://www.pnas.org/content/108/30/12233.full.pdf} \BibitemShut {NoStop}%
\bibitem [{\citenamefont {Cao}\ \emph {et~al.}(2018{\natexlab{a}})\citenamefont
  {Cao}, \citenamefont {Fatemi}, \citenamefont {Demir}, \citenamefont {Fang},
  \citenamefont {Tomarken}, \citenamefont {Luo}, \citenamefont
  {Sanchez-Yamagishi}, \citenamefont {Watanabe}, \citenamefont {Taniguchi},
  \citenamefont {Kaxiras}, \citenamefont {Ashoori},\ and\ \citenamefont
  {Jarillo-Herrero}}]{CaoCorrelatedInsulator}%
  \BibitemOpen
  \bibfield  {author} {\bibinfo {author} {\bibfnamefont {Y.}~\bibnamefont
  {Cao}}, \bibinfo {author} {\bibfnamefont {V.}~\bibnamefont {Fatemi}},
  \bibinfo {author} {\bibfnamefont {A.}~\bibnamefont {Demir}}, \bibinfo
  {author} {\bibfnamefont {S.}~\bibnamefont {Fang}}, \bibinfo {author}
  {\bibfnamefont {S.~L.}\ \bibnamefont {Tomarken}}, \bibinfo {author}
  {\bibfnamefont {J.~Y.}\ \bibnamefont {Luo}}, \bibinfo {author} {\bibfnamefont
  {J.~D.}\ \bibnamefont {Sanchez-Yamagishi}}, \bibinfo {author} {\bibfnamefont
  {K.}~\bibnamefont {Watanabe}}, \bibinfo {author} {\bibfnamefont
  {T.}~\bibnamefont {Taniguchi}}, \bibinfo {author} {\bibfnamefont
  {E.}~\bibnamefont {Kaxiras}}, \bibinfo {author} {\bibfnamefont {R.~C.}\
  \bibnamefont {Ashoori}}, \ and\ \bibinfo {author} {\bibfnamefont
  {P.}~\bibnamefont {Jarillo-Herrero}},\ }\href {\doibase 10.1038/nature26154}
  {\bibfield  {journal} {\bibinfo  {journal} {Nature}\ }\textbf {\bibinfo
  {volume} {556}},\ \bibinfo {pages} {80} (\bibinfo {year}
  {2018}{\natexlab{a}})}\BibitemShut {NoStop}%
\bibitem [{\citenamefont {Cao}\ \emph {et~al.}(2018{\natexlab{b}})\citenamefont
  {Cao}, \citenamefont {Fatemi}, \citenamefont {Fang}, \citenamefont
  {Watanabe}, \citenamefont {Taniguchi}, \citenamefont {Kaxiras},\ and\
  \citenamefont {Jarillo-Herrero}}]{CaoUnconventionalSC}%
  \BibitemOpen
  \bibfield  {author} {\bibinfo {author} {\bibfnamefont {Y.}~\bibnamefont
  {Cao}}, \bibinfo {author} {\bibfnamefont {V.}~\bibnamefont {Fatemi}},
  \bibinfo {author} {\bibfnamefont {S.}~\bibnamefont {Fang}}, \bibinfo {author}
  {\bibfnamefont {K.}~\bibnamefont {Watanabe}}, \bibinfo {author}
  {\bibfnamefont {T.}~\bibnamefont {Taniguchi}}, \bibinfo {author}
  {\bibfnamefont {E.}~\bibnamefont {Kaxiras}}, \ and\ \bibinfo {author}
  {\bibfnamefont {P.}~\bibnamefont {Jarillo-Herrero}},\ }\href {\doibase
  10.1038/nature26160} {\bibfield  {journal} {\bibinfo  {journal} {Nature}\
  }\textbf {\bibinfo {volume} {556}},\ \bibinfo {pages} {43} (\bibinfo {year}
  {2018}{\natexlab{b}})}\BibitemShut {NoStop}%
\bibitem [{\citenamefont {Yankowitz}\ \emph {et~al.}(2019)\citenamefont
  {Yankowitz}, \citenamefont {Chen}, \citenamefont {Polshyn}, \citenamefont
  {Zhang}, \citenamefont {Watanabe}, \citenamefont {Taniguchi}, \citenamefont
  {Graf}, \citenamefont {Young},\ and\ \citenamefont
  {Dean}}]{YankowitzTuningMATBG}%
  \BibitemOpen
  \bibfield  {author} {\bibinfo {author} {\bibfnamefont {M.}~\bibnamefont
  {Yankowitz}}, \bibinfo {author} {\bibfnamefont {S.}~\bibnamefont {Chen}},
  \bibinfo {author} {\bibfnamefont {H.}~\bibnamefont {Polshyn}}, \bibinfo
  {author} {\bibfnamefont {Y.}~\bibnamefont {Zhang}}, \bibinfo {author}
  {\bibfnamefont {K.}~\bibnamefont {Watanabe}}, \bibinfo {author}
  {\bibfnamefont {T.}~\bibnamefont {Taniguchi}}, \bibinfo {author}
  {\bibfnamefont {D.}~\bibnamefont {Graf}}, \bibinfo {author} {\bibfnamefont
  {A.~F.}\ \bibnamefont {Young}}, \ and\ \bibinfo {author} {\bibfnamefont
  {C.~R.}\ \bibnamefont {Dean}},\ }\href {\doibase 10.1126/science.aav1910}
  {\bibfield  {journal} {\bibinfo  {journal} {Science}\ }\textbf {\bibinfo
  {volume} {363}},\ \bibinfo {pages} {1059} (\bibinfo {year} {2019})},\ \Eprint
  {http://arxiv.org/abs/https://science.sciencemag.org/content/363/6431/1059.full.pdf}
  {https://science.sciencemag.org/content/363/6431/1059.full.pdf} \BibitemShut
  {NoStop}%
\bibitem [{\citenamefont {Saito}\ \emph
  {et~al.}(2020{\natexlab{a}})\citenamefont {Saito}, \citenamefont {Ge},
  \citenamefont {Watanabe}, \citenamefont {Taniguchi},\ and\ \citenamefont
  {Young}}]{YoungTuningSC}%
  \BibitemOpen
  \bibfield  {author} {\bibinfo {author} {\bibfnamefont {Y.}~\bibnamefont
  {Saito}}, \bibinfo {author} {\bibfnamefont {J.}~\bibnamefont {Ge}}, \bibinfo
  {author} {\bibfnamefont {K.}~\bibnamefont {Watanabe}}, \bibinfo {author}
  {\bibfnamefont {T.}~\bibnamefont {Taniguchi}}, \ and\ \bibinfo {author}
  {\bibfnamefont {A.~F.}\ \bibnamefont {Young}},\ }\href {\doibase
  10.1038/s41567-020-0928-3} {\bibfield  {journal} {\bibinfo  {journal} {Nature
  Physics}\ }\textbf {\bibinfo {volume} {16}},\ \bibinfo {pages} {926}
  (\bibinfo {year} {2020}{\natexlab{a}})}\BibitemShut {NoStop}%
\bibitem [{\citenamefont {Lu}\ \emph {et~al.}(2019)\citenamefont {Lu},
  \citenamefont {Stepanov}, \citenamefont {Yang}, \citenamefont {Xie},
  \citenamefont {Aamir}, \citenamefont {Das}, \citenamefont {Urgell},
  \citenamefont {Watanabe}, \citenamefont {Taniguchi}, \citenamefont {Zhang},
  \citenamefont {Bachtold}, \citenamefont {MacDonald},\ and\ \citenamefont
  {Efetov}}]{EfetovAllIntegers}%
  \BibitemOpen
  \bibfield  {author} {\bibinfo {author} {\bibfnamefont {X.}~\bibnamefont
  {Lu}}, \bibinfo {author} {\bibfnamefont {P.}~\bibnamefont {Stepanov}},
  \bibinfo {author} {\bibfnamefont {W.}~\bibnamefont {Yang}}, \bibinfo {author}
  {\bibfnamefont {M.}~\bibnamefont {Xie}}, \bibinfo {author} {\bibfnamefont
  {M.~A.}\ \bibnamefont {Aamir}}, \bibinfo {author} {\bibfnamefont
  {I.}~\bibnamefont {Das}}, \bibinfo {author} {\bibfnamefont {C.}~\bibnamefont
  {Urgell}}, \bibinfo {author} {\bibfnamefont {K.}~\bibnamefont {Watanabe}},
  \bibinfo {author} {\bibfnamefont {T.}~\bibnamefont {Taniguchi}}, \bibinfo
  {author} {\bibfnamefont {G.}~\bibnamefont {Zhang}}, \bibinfo {author}
  {\bibfnamefont {A.}~\bibnamefont {Bachtold}}, \bibinfo {author}
  {\bibfnamefont {A.~H.}\ \bibnamefont {MacDonald}}, \ and\ \bibinfo {author}
  {\bibfnamefont {D.~K.}\ \bibnamefont {Efetov}},\ }\href {\doibase
  10.1038/s41586-019-1695-0} {\bibfield  {journal} {\bibinfo  {journal}
  {Nature}\ }\textbf {\bibinfo {volume} {574}},\ \bibinfo {pages} {653}
  (\bibinfo {year} {2019})}\BibitemShut {NoStop}%
\bibitem [{\citenamefont {Po}\ \emph {et~al.}(2018)\citenamefont {Po},
  \citenamefont {Zou}, \citenamefont {Vishwanath},\ and\ \citenamefont
  {Senthil}}]{SenthilAshvin2018PRX}%
  \BibitemOpen
  \bibfield  {author} {\bibinfo {author} {\bibfnamefont {H.~C.}\ \bibnamefont
  {Po}}, \bibinfo {author} {\bibfnamefont {L.}~\bibnamefont {Zou}}, \bibinfo
  {author} {\bibfnamefont {A.}~\bibnamefont {Vishwanath}}, \ and\ \bibinfo
  {author} {\bibfnamefont {T.}~\bibnamefont {Senthil}},\ }\href {\doibase
  10.1103/PhysRevX.8.031089} {\bibfield  {journal} {\bibinfo  {journal} {Phys.
  Rev. X}\ }\textbf {\bibinfo {volume} {8}},\ \bibinfo {pages} {031089}
  (\bibinfo {year} {2018})}\BibitemShut {NoStop}%
\bibitem [{\citenamefont {Dodaro}\ \emph {et~al.}(2018)\citenamefont {Dodaro},
  \citenamefont {Kivelson}, \citenamefont {Schattner}, \citenamefont {Sun},\
  and\ \citenamefont {Wang}}]{Dodaro2018}%
  \BibitemOpen
  \bibfield  {author} {\bibinfo {author} {\bibfnamefont {J.~F.}\ \bibnamefont
  {Dodaro}}, \bibinfo {author} {\bibfnamefont {S.~A.}\ \bibnamefont
  {Kivelson}}, \bibinfo {author} {\bibfnamefont {Y.}~\bibnamefont {Schattner}},
  \bibinfo {author} {\bibfnamefont {X.~Q.}\ \bibnamefont {Sun}}, \ and\
  \bibinfo {author} {\bibfnamefont {C.}~\bibnamefont {Wang}},\ }\href {\doibase
  10.1103/PhysRevB.98.075154} {\bibfield  {journal} {\bibinfo  {journal} {Phys.
  Rev. B}\ }\textbf {\bibinfo {volume} {98}},\ \bibinfo {pages} {075154}
  (\bibinfo {year} {2018})}\BibitemShut {NoStop}%
\bibitem [{\citenamefont {Kang}\ and\ \citenamefont {Vafek}(2019)}]{Kang2019}%
  \BibitemOpen
  \bibfield  {author} {\bibinfo {author} {\bibfnamefont {J.}~\bibnamefont
  {Kang}}\ and\ \bibinfo {author} {\bibfnamefont {O.}~\bibnamefont {Vafek}},\
  }\href {\doibase 10.1103/PhysRevLett.122.246401} {\bibfield  {journal}
  {\bibinfo  {journal} {Phys. Rev. Lett.}\ }\textbf {\bibinfo {volume} {122}},\
  \bibinfo {pages} {246401} (\bibinfo {year} {2019})}\BibitemShut {NoStop}%
\bibitem [{\citenamefont {Seo}\ \emph {et~al.}(2019)\citenamefont {Seo},
  \citenamefont {Kotov},\ and\ \citenamefont {Uchoa}}]{UchoaQuarterfilling}%
  \BibitemOpen
  \bibfield  {author} {\bibinfo {author} {\bibfnamefont {K.}~\bibnamefont
  {Seo}}, \bibinfo {author} {\bibfnamefont {V.~N.}\ \bibnamefont {Kotov}}, \
  and\ \bibinfo {author} {\bibfnamefont {B.}~\bibnamefont {Uchoa}},\ }\href
  {\doibase 10.1103/PhysRevLett.122.246402} {\bibfield  {journal} {\bibinfo
  {journal} {Phys. Rev. Lett.}\ }\textbf {\bibinfo {volume} {122}},\ \bibinfo
  {pages} {246402} (\bibinfo {year} {2019})}\BibitemShut {NoStop}%
\bibitem [{\citenamefont {Bultinck}\ \emph
  {et~al.}(2020{\natexlab{a}})\citenamefont {Bultinck}, \citenamefont {Khalaf},
  \citenamefont {Liu}, \citenamefont {Chatterjee}, \citenamefont {Vishwanath},\
  and\ \citenamefont {Zaletel}}]{BultnickKhalaf2020}%
  \BibitemOpen
  \bibfield  {author} {\bibinfo {author} {\bibfnamefont {N.}~\bibnamefont
  {Bultinck}}, \bibinfo {author} {\bibfnamefont {E.}~\bibnamefont {Khalaf}},
  \bibinfo {author} {\bibfnamefont {S.}~\bibnamefont {Liu}}, \bibinfo {author}
  {\bibfnamefont {S.}~\bibnamefont {Chatterjee}}, \bibinfo {author}
  {\bibfnamefont {A.}~\bibnamefont {Vishwanath}}, \ and\ \bibinfo {author}
  {\bibfnamefont {M.~P.}\ \bibnamefont {Zaletel}},\ }\href {\doibase
  10.1103/PhysRevX.10.031034} {\bibfield  {journal} {\bibinfo  {journal} {Phys.
  Rev. X}\ }\textbf {\bibinfo {volume} {10}},\ \bibinfo {pages} {031034}
  (\bibinfo {year} {2020}{\natexlab{a}})}\BibitemShut {NoStop}%
\bibitem [{\citenamefont {Kang}\ and\ \citenamefont {Vafek}(2020)}]{Kang2020}%
  \BibitemOpen
  \bibfield  {author} {\bibinfo {author} {\bibfnamefont {J.}~\bibnamefont
  {Kang}}\ and\ \bibinfo {author} {\bibfnamefont {O.}~\bibnamefont {Vafek}},\
  }\href {\doibase 10.1103/PhysRevB.102.035161} {\bibfield  {journal} {\bibinfo
   {journal} {Phys. Rev. B}\ }\textbf {\bibinfo {volume} {102}},\ \bibinfo
  {pages} {035161} (\bibinfo {year} {2020})}\BibitemShut {NoStop}%
\bibitem [{\citenamefont {Vafek}\ and\ \citenamefont {Kang}(2020)}]{VAFEKRG}%
  \BibitemOpen
  \bibfield  {author} {\bibinfo {author} {\bibfnamefont {O.}~\bibnamefont
  {Vafek}}\ and\ \bibinfo {author} {\bibfnamefont {J.}~\bibnamefont {Kang}},\
  }\href {\doibase 10.1103/PhysRevLett.125.257602} {\bibfield  {journal}
  {\bibinfo  {journal} {Phys. Rev. Lett.}\ }\textbf {\bibinfo {volume} {125}},\
  \bibinfo {pages} {257602} (\bibinfo {year} {2020})}\BibitemShut {NoStop}%
\bibitem [{\citenamefont {Lian}\ \emph {et~al.}(2021)\citenamefont {Lian},
  \citenamefont {Song}, \citenamefont {Regnault}, \citenamefont {Efetov},
  \citenamefont {Yazdani},\ and\ \citenamefont
  {Bernevig}}]{TBG4Exactinsulatorgroundphasediagram}%
  \BibitemOpen
  \bibfield  {author} {\bibinfo {author} {\bibfnamefont {B.}~\bibnamefont
  {Lian}}, \bibinfo {author} {\bibfnamefont {Z.-D.}\ \bibnamefont {Song}},
  \bibinfo {author} {\bibfnamefont {N.}~\bibnamefont {Regnault}}, \bibinfo
  {author} {\bibfnamefont {D.~K.}\ \bibnamefont {Efetov}}, \bibinfo {author}
  {\bibfnamefont {A.}~\bibnamefont {Yazdani}}, \ and\ \bibinfo {author}
  {\bibfnamefont {B.~A.}\ \bibnamefont {Bernevig}},\ }\href {\doibase
  10.1103/PhysRevB.103.205414} {\bibfield  {journal} {\bibinfo  {journal}
  {Phys. Rev. B}\ }\textbf {\bibinfo {volume} {103}},\ \bibinfo {pages}
  {205414} (\bibinfo {year} {2021})}\BibitemShut {NoStop}%
\bibitem [{\citenamefont {Xie}\ \emph {et~al.}(2021)\citenamefont {Xie},
  \citenamefont {Cowsik}, \citenamefont {Song}, \citenamefont {Lian},
  \citenamefont {Bernevig},\ and\ \citenamefont
  {Regnault}}]{TBG6exactdiagonalizationintegerfillings}%
  \BibitemOpen
  \bibfield  {author} {\bibinfo {author} {\bibfnamefont {F.}~\bibnamefont
  {Xie}}, \bibinfo {author} {\bibfnamefont {A.}~\bibnamefont {Cowsik}},
  \bibinfo {author} {\bibfnamefont {Z.-D.}\ \bibnamefont {Song}}, \bibinfo
  {author} {\bibfnamefont {B.}~\bibnamefont {Lian}}, \bibinfo {author}
  {\bibfnamefont {B.~A.}\ \bibnamefont {Bernevig}}, \ and\ \bibinfo {author}
  {\bibfnamefont {N.}~\bibnamefont {Regnault}},\ }\href {\doibase
  10.1103/PhysRevB.103.205416} {\bibfield  {journal} {\bibinfo  {journal}
  {Phys. Rev. B}\ }\textbf {\bibinfo {volume} {103}},\ \bibinfo {pages}
  {205416} (\bibinfo {year} {2021})}\BibitemShut {NoStop}%
\bibitem [{\citenamefont {Stepanov}\ \emph {et~al.}(2020)\citenamefont
  {Stepanov}, \citenamefont {Das}, \citenamefont {Lu}, \citenamefont
  {Fahimniya}, \citenamefont {Watanabe}, \citenamefont {Taniguchi},
  \citenamefont {Koppens}, \citenamefont {Lischner}, \citenamefont {Levitov},\
  and\ \citenamefont {Efetov}}]{EfetovTuningSC}%
  \BibitemOpen
  \bibfield  {author} {\bibinfo {author} {\bibfnamefont {P.}~\bibnamefont
  {Stepanov}}, \bibinfo {author} {\bibfnamefont {I.}~\bibnamefont {Das}},
  \bibinfo {author} {\bibfnamefont {X.}~\bibnamefont {Lu}}, \bibinfo {author}
  {\bibfnamefont {A.}~\bibnamefont {Fahimniya}}, \bibinfo {author}
  {\bibfnamefont {K.}~\bibnamefont {Watanabe}}, \bibinfo {author}
  {\bibfnamefont {T.}~\bibnamefont {Taniguchi}}, \bibinfo {author}
  {\bibfnamefont {F.~H.~L.}\ \bibnamefont {Koppens}}, \bibinfo {author}
  {\bibfnamefont {J.}~\bibnamefont {Lischner}}, \bibinfo {author}
  {\bibfnamefont {L.}~\bibnamefont {Levitov}}, \ and\ \bibinfo {author}
  {\bibfnamefont {D.~K.}\ \bibnamefont {Efetov}},\ }\href {\doibase
  10.1038/s41586-020-2459-6} {\bibfield  {journal} {\bibinfo  {journal}
  {Nature}\ }\textbf {\bibinfo {volume} {583}},\ \bibinfo {pages} {375}
  (\bibinfo {year} {2020})}\BibitemShut {NoStop}%
\bibitem [{\citenamefont {Liu}\ \emph {et~al.}(2020)\citenamefont {Liu},
  \citenamefont {Wang}, \citenamefont {Watanabe}, \citenamefont {Taniguchi},
  \citenamefont {Vafek},\ and\ \citenamefont {Li}}]{BLGscreening}%
  \BibitemOpen
  \bibfield  {author} {\bibinfo {author} {\bibfnamefont {X.}~\bibnamefont
  {Liu}}, \bibinfo {author} {\bibfnamefont {Z.}~\bibnamefont {Wang}}, \bibinfo
  {author} {\bibfnamefont {K.}~\bibnamefont {Watanabe}}, \bibinfo {author}
  {\bibfnamefont {T.}~\bibnamefont {Taniguchi}}, \bibinfo {author}
  {\bibfnamefont {O.}~\bibnamefont {Vafek}}, \ and\ \bibinfo {author}
  {\bibfnamefont {J.~I.~A.}\ \bibnamefont {Li}},\ }\href@noop {} {\enquote
  {\bibinfo {title} {Tuning electron correlation in magic-angle twisted bilayer
  graphene using coulomb screening},}\ } (\bibinfo {year} {2020}),\ \Eprint
  {http://arxiv.org/abs/arXiv:2003.11072} {arXiv:2003.11072} \BibitemShut
  {NoStop}%
\bibitem [{\citenamefont {Wu}\ \emph {et~al.}(2018)\citenamefont {Wu},
  \citenamefont {MacDonald},\ and\ \citenamefont
  {Martin}}]{MacdonaldPRLopticalPhonons}%
  \BibitemOpen
  \bibfield  {author} {\bibinfo {author} {\bibfnamefont {F.}~\bibnamefont
  {Wu}}, \bibinfo {author} {\bibfnamefont {A.~H.}\ \bibnamefont {MacDonald}}, \
  and\ \bibinfo {author} {\bibfnamefont {I.}~\bibnamefont {Martin}},\ }\href
  {\doibase 10.1103/PhysRevLett.121.257001} {\bibfield  {journal} {\bibinfo
  {journal} {Phys. Rev. Lett.}\ }\textbf {\bibinfo {volume} {121}},\ \bibinfo
  {pages} {257001} (\bibinfo {year} {2018})}\BibitemShut {NoStop}%
\bibitem [{\citenamefont {Lian}\ \emph {et~al.}(2019)\citenamefont {Lian},
  \citenamefont {Wang},\ and\ \citenamefont
  {Bernevig}}]{LianBioBernevigPRLphonons}%
  \BibitemOpen
  \bibfield  {author} {\bibinfo {author} {\bibfnamefont {B.}~\bibnamefont
  {Lian}}, \bibinfo {author} {\bibfnamefont {Z.}~\bibnamefont {Wang}}, \ and\
  \bibinfo {author} {\bibfnamefont {B.~A.}\ \bibnamefont {Bernevig}},\ }\href
  {\doibase 10.1103/PhysRevLett.122.257002} {\bibfield  {journal} {\bibinfo
  {journal} {Phys. Rev. Lett.}\ }\textbf {\bibinfo {volume} {122}},\ \bibinfo
  {pages} {257002} (\bibinfo {year} {2019})}\BibitemShut {NoStop}%
\bibitem [{\citenamefont {Wu}\ \emph {et~al.}(2019)\citenamefont {Wu},
  \citenamefont {Hwang},\ and\ \citenamefont {Das~Sarma}}]{Wu2019}%
  \BibitemOpen
  \bibfield  {author} {\bibinfo {author} {\bibfnamefont {F.}~\bibnamefont
  {Wu}}, \bibinfo {author} {\bibfnamefont {E.}~\bibnamefont {Hwang}}, \ and\
  \bibinfo {author} {\bibfnamefont {S.}~\bibnamefont {Das~Sarma}},\ }\href
  {\doibase 10.1103/PhysRevB.99.165112} {\bibfield  {journal} {\bibinfo
  {journal} {Phys. Rev. B}\ }\textbf {\bibinfo {volume} {99}},\ \bibinfo
  {pages} {165112} (\bibinfo {year} {2019})}\BibitemShut {NoStop}%
\bibitem [{\citenamefont {Bernevig}\ \emph {et~al.}(2021)\citenamefont
  {Bernevig}, \citenamefont {Lian}, \citenamefont {Cowsik}, \citenamefont
  {Xie}, \citenamefont {Regnault},\ and\ \citenamefont
  {Song}}]{TBG5phononsNotCoulomb}%
  \BibitemOpen
  \bibfield  {author} {\bibinfo {author} {\bibfnamefont {B.~A.}\ \bibnamefont
  {Bernevig}}, \bibinfo {author} {\bibfnamefont {B.}~\bibnamefont {Lian}},
  \bibinfo {author} {\bibfnamefont {A.}~\bibnamefont {Cowsik}}, \bibinfo
  {author} {\bibfnamefont {F.}~\bibnamefont {Xie}}, \bibinfo {author}
  {\bibfnamefont {N.}~\bibnamefont {Regnault}}, \ and\ \bibinfo {author}
  {\bibfnamefont {Z.-D.}\ \bibnamefont {Song}},\ }\href {\doibase
  10.1103/PhysRevB.103.205415} {\bibfield  {journal} {\bibinfo  {journal}
  {Phys. Rev. B}\ }\textbf {\bibinfo {volume} {103}},\ \bibinfo {pages}
  {205415} (\bibinfo {year} {2021})}\BibitemShut {NoStop}%
\bibitem [{\citenamefont {Lewandowski}\ \emph
  {et~al.}(2021{\natexlab{a}})\citenamefont {Lewandowski}, \citenamefont
  {Chowdhury},\ and\ \citenamefont {Ruhman}}]{Ruhman2021}%
  \BibitemOpen
  \bibfield  {author} {\bibinfo {author} {\bibfnamefont {C.}~\bibnamefont
  {Lewandowski}}, \bibinfo {author} {\bibfnamefont {D.}~\bibnamefont
  {Chowdhury}}, \ and\ \bibinfo {author} {\bibfnamefont {J.}~\bibnamefont
  {Ruhman}},\ }\href {\doibase 10.1103/PhysRevB.103.235401} {\bibfield
  {journal} {\bibinfo  {journal} {Phys. Rev. B}\ }\textbf {\bibinfo {volume}
  {103}},\ \bibinfo {pages} {235401} (\bibinfo {year}
  {2021}{\natexlab{a}})}\BibitemShut {NoStop}%
\bibitem [{\citenamefont {Wong}\ \emph {et~al.}(2020)\citenamefont {Wong},
  \citenamefont {Nuckolls}, \citenamefont {Oh}, \citenamefont {Lian},
  \citenamefont {Xie}, \citenamefont {Jeon}, \citenamefont {Watanabe},
  \citenamefont {Taniguchi}, \citenamefont {Bernevig},\ and\ \citenamefont
  {Yazdani}}]{YazdaniRevivals}%
  \BibitemOpen
  \bibfield  {author} {\bibinfo {author} {\bibfnamefont {D.}~\bibnamefont
  {Wong}}, \bibinfo {author} {\bibfnamefont {K.~P.}\ \bibnamefont {Nuckolls}},
  \bibinfo {author} {\bibfnamefont {M.}~\bibnamefont {Oh}}, \bibinfo {author}
  {\bibfnamefont {B.}~\bibnamefont {Lian}}, \bibinfo {author} {\bibfnamefont
  {Y.}~\bibnamefont {Xie}}, \bibinfo {author} {\bibfnamefont {S.}~\bibnamefont
  {Jeon}}, \bibinfo {author} {\bibfnamefont {K.}~\bibnamefont {Watanabe}},
  \bibinfo {author} {\bibfnamefont {T.}~\bibnamefont {Taniguchi}}, \bibinfo
  {author} {\bibfnamefont {B.~A.}\ \bibnamefont {Bernevig}}, \ and\ \bibinfo
  {author} {\bibfnamefont {A.}~\bibnamefont {Yazdani}},\ }\href {\doibase
  10.1038/s41586-020-2339-0} {\bibfield  {journal} {\bibinfo  {journal}
  {Nature}\ }\textbf {\bibinfo {volume} {582}},\ \bibinfo {pages} {198}
  (\bibinfo {year} {2020})}\BibitemShut {NoStop}%
\bibitem [{\citenamefont {Zondiner}\ \emph {et~al.}(2020)\citenamefont
  {Zondiner}, \citenamefont {Rozen}, \citenamefont {Rodan-Legrain},
  \citenamefont {Cao}, \citenamefont {Queiroz}, \citenamefont {Taniguchi},
  \citenamefont {Watanabe}, \citenamefont {Oreg}, \citenamefont {von Oppen},
  \citenamefont {Stern}, \citenamefont {Berg}, \citenamefont
  {Jarillo-Herrero},\ and\ \citenamefont {Ilani}}]{DiracRevivals}%
  \BibitemOpen
  \bibfield  {author} {\bibinfo {author} {\bibfnamefont {U.}~\bibnamefont
  {Zondiner}}, \bibinfo {author} {\bibfnamefont {A.}~\bibnamefont {Rozen}},
  \bibinfo {author} {\bibfnamefont {D.}~\bibnamefont {Rodan-Legrain}}, \bibinfo
  {author} {\bibfnamefont {Y.}~\bibnamefont {Cao}}, \bibinfo {author}
  {\bibfnamefont {R.}~\bibnamefont {Queiroz}}, \bibinfo {author} {\bibfnamefont
  {T.}~\bibnamefont {Taniguchi}}, \bibinfo {author} {\bibfnamefont
  {K.}~\bibnamefont {Watanabe}}, \bibinfo {author} {\bibfnamefont
  {Y.}~\bibnamefont {Oreg}}, \bibinfo {author} {\bibfnamefont {F.}~\bibnamefont
  {von Oppen}}, \bibinfo {author} {\bibfnamefont {A.}~\bibnamefont {Stern}},
  \bibinfo {author} {\bibfnamefont {E.}~\bibnamefont {Berg}}, \bibinfo {author}
  {\bibfnamefont {P.}~\bibnamefont {Jarillo-Herrero}}, \ and\ \bibinfo {author}
  {\bibfnamefont {S.}~\bibnamefont {Ilani}},\ }\href {\doibase
  10.1038/s41586-020-2373-y} {\bibfield  {journal} {\bibinfo  {journal}
  {Nature}\ }\textbf {\bibinfo {volume} {582}},\ \bibinfo {pages} {203}
  (\bibinfo {year} {2020})}\BibitemShut {NoStop}%
\bibitem [{\citenamefont {Kang}\ \emph {et~al.}(2021)\citenamefont {Kang},
  \citenamefont {Bernevig},\ and\ \citenamefont
  {Vafek}}]{kang2021cascadesVafekBernevig}%
  \BibitemOpen
  \bibfield  {author} {\bibinfo {author} {\bibfnamefont {J.}~\bibnamefont
  {Kang}}, \bibinfo {author} {\bibfnamefont {B.~A.}\ \bibnamefont {Bernevig}},
  \ and\ \bibinfo {author} {\bibfnamefont {O.}~\bibnamefont {Vafek}},\
  }\href@noop {} {\enquote {\bibinfo {title} {Cascades between light and heavy
  fermions in the normal state of magic angle twisted bilayer graphene},}\ }
  (\bibinfo {year} {2021}),\ \Eprint {http://arxiv.org/abs/2104.01145}
  {arXiv:2104.01145 [cond-mat.str-el]} \BibitemShut {NoStop}%
\bibitem [{C3h()}]{C3h0}%
  \BibitemOpen
  \href@noop {} {}\bibinfo {note} {The single-particle term in the Hamiltonian
  we use is not $C_3$-symmetric, unlike the MATBG dispersion. However, within
  this model it has little consequence, as we also do not explore spontaneous
  $C_3$ symmetry breaking.}\BibitemShut {Stop}%
\bibitem [{Sup()}]{SupplementRef}%
  \BibitemOpen
  \href@noop {} {}\bibinfo {note} {See Supplemental Material for details
  regarding the single-particle Hamiltonian, the variational mean-field
  approach, calculation of $\tilde{\nu}$, the superconducting RG equation, the
  role of superconducting phase fluctuations, the effect of electrostatic
  screening, estimation of relevant coupling constants, and suppression of
  superconductivity by alignment with h-BN, which includes
  Refs.~\cite{AllenDynesLinearinLambdaTc,GrapheneBCS,abrikosov2017fundamentals,ASLAMASOV1968238,hBNgapsize,VortexLattice,maki1969gapless}}\BibitemShut
  {NoStop}%
\bibitem [{che()}]{check}%
  \BibitemOpen
  \href@noop {} {}\bibinfo {note} {It can be directly checked for example that
  $(\Psi^\dagger C_3 \vec{\cal O}_\alpha C_3^\dagger \Psi) ^2 = (\Psi^\dagger
  \vec{\cal O}_\alpha \Psi)$ for $\alpha=1,\dots, 4$.}\BibitemShut {Stop}%
\bibitem [{QHF()}]{QHFM}%
  \BibitemOpen
  \href@noop {} {}\bibinfo {note} {This sensitivity is a common feature in
  quantum Hall ferromagnets \cite{MacdonaldQHFM,KharitonovGraphene}, whose
  physics which is arguably pertinent to the theory of MATBG.}\BibitemShut
  {Stop}%
\bibitem [{\citenamefont {Hofmann}\ \emph {et~al.}(2021)\citenamefont
  {Hofmann}, \citenamefont {Khalaf}, \citenamefont {Vishwanath}, \citenamefont
  {Berg},\ and\ \citenamefont {Lee}}]{hofmann2021fermionic}%
  \BibitemOpen
  \bibfield  {author} {\bibinfo {author} {\bibfnamefont {J.~S.}\ \bibnamefont
  {Hofmann}}, \bibinfo {author} {\bibfnamefont {E.}~\bibnamefont {Khalaf}},
  \bibinfo {author} {\bibfnamefont {A.}~\bibnamefont {Vishwanath}}, \bibinfo
  {author} {\bibfnamefont {E.}~\bibnamefont {Berg}}, \ and\ \bibinfo {author}
  {\bibfnamefont {J.~Y.}\ \bibnamefont {Lee}},\ }\href@noop {} {\enquote
  {\bibinfo {title} {Fermionic monte carlo study of a realistic model of
  twisted bilayer graphene},}\ } (\bibinfo {year} {2021}),\ \Eprint
  {http://arxiv.org/abs/2105.12112} {arXiv:2105.12112 [cond-mat.str-el]}
  \BibitemShut {NoStop}%
\bibitem [{\citenamefont {Zhou}\ \emph {et~al.}(2016)\citenamefont {Zhou},
  \citenamefont {Yuan}, \citenamefont {Jiang},\ and\ \citenamefont
  {Law}}]{KTLawPairingCorrelationsTriplet}%
  \BibitemOpen
  \bibfield  {author} {\bibinfo {author} {\bibfnamefont {B.~T.}\ \bibnamefont
  {Zhou}}, \bibinfo {author} {\bibfnamefont {N.~F.~Q.}\ \bibnamefont {Yuan}},
  \bibinfo {author} {\bibfnamefont {H.-L.}\ \bibnamefont {Jiang}}, \ and\
  \bibinfo {author} {\bibfnamefont {K.~T.}\ \bibnamefont {Law}},\ }\href
  {\doibase 10.1103/PhysRevB.93.180501} {\bibfield  {journal} {\bibinfo
  {journal} {Phys. Rev. B}\ }\textbf {\bibinfo {volume} {93}},\ \bibinfo
  {pages} {180501} (\bibinfo {year} {2016})}\BibitemShut {NoStop}%
\bibitem [{\citenamefont {Clogston}(1962)}]{ClogsonLimit}%
  \BibitemOpen
  \bibfield  {author} {\bibinfo {author} {\bibfnamefont {A.~M.}\ \bibnamefont
  {Clogston}},\ }\href {\doibase 10.1103/PhysRevLett.9.266} {\bibfield
  {journal} {\bibinfo  {journal} {Phys. Rev. Lett.}\ }\textbf {\bibinfo
  {volume} {9}},\ \bibinfo {pages} {266} (\bibinfo {year} {1962})}\BibitemShut
  {NoStop}%
\bibitem [{\citenamefont {Lu}\ \emph {et~al.}(2015)\citenamefont {Lu},
  \citenamefont {Zheliuk}, \citenamefont {Leermakers}, \citenamefont {Yuan},
  \citenamefont {Zeitler}, \citenamefont {Law},\ and\ \citenamefont
  {Ye}}]{IsingSuperconductivityMoS2}%
  \BibitemOpen
  \bibfield  {author} {\bibinfo {author} {\bibfnamefont {J.~M.}\ \bibnamefont
  {Lu}}, \bibinfo {author} {\bibfnamefont {O.}~\bibnamefont {Zheliuk}},
  \bibinfo {author} {\bibfnamefont {I.}~\bibnamefont {Leermakers}}, \bibinfo
  {author} {\bibfnamefont {N.~F.~Q.}\ \bibnamefont {Yuan}}, \bibinfo {author}
  {\bibfnamefont {U.}~\bibnamefont {Zeitler}}, \bibinfo {author} {\bibfnamefont
  {K.~T.}\ \bibnamefont {Law}}, \ and\ \bibinfo {author} {\bibfnamefont
  {J.~T.}\ \bibnamefont {Ye}},\ }\href {\doibase 10.1126/science.aab2277}
  {\bibfield  {journal} {\bibinfo  {journal} {Science}\ }\textbf {\bibinfo
  {volume} {350}},\ \bibinfo {pages} {1353} (\bibinfo {year} {2015})},\ \Eprint
  {http://arxiv.org/abs/https://science.sciencemag.org/content/350/6266/1353.full.pdf}
  {https://science.sciencemag.org/content/350/6266/1353.full.pdf} \BibitemShut
  {NoStop}%
\bibitem [{\citenamefont {Xi}\ \emph {et~al.}(2016)\citenamefont {Xi},
  \citenamefont {Wang}, \citenamefont {Zhao}, \citenamefont {Park},
  \citenamefont {Law}, \citenamefont {Berger}, \citenamefont {Forr{\'o}},
  \citenamefont {Shan},\ and\ \citenamefont
  {Mak}}]{IsingSuperconductivityNbSe2}%
  \BibitemOpen
  \bibfield  {author} {\bibinfo {author} {\bibfnamefont {X.}~\bibnamefont
  {Xi}}, \bibinfo {author} {\bibfnamefont {Z.}~\bibnamefont {Wang}}, \bibinfo
  {author} {\bibfnamefont {W.}~\bibnamefont {Zhao}}, \bibinfo {author}
  {\bibfnamefont {J.-H.}\ \bibnamefont {Park}}, \bibinfo {author}
  {\bibfnamefont {K.~T.}\ \bibnamefont {Law}}, \bibinfo {author} {\bibfnamefont
  {H.}~\bibnamefont {Berger}}, \bibinfo {author} {\bibfnamefont
  {L.}~\bibnamefont {Forr{\'o}}}, \bibinfo {author} {\bibfnamefont
  {J.}~\bibnamefont {Shan}}, \ and\ \bibinfo {author} {\bibfnamefont {K.~F.}\
  \bibnamefont {Mak}},\ }\href {\doibase 10.1038/nphys3538} {\bibfield
  {journal} {\bibinfo  {journal} {Nature Physics}\ }\textbf {\bibinfo {volume}
  {12}},\ \bibinfo {pages} {139} (\bibinfo {year} {2016})}\BibitemShut
  {NoStop}%
\bibitem [{\citenamefont {Tolmachev}(1962)}]{tolmachev1962logarithmic}%
  \BibitemOpen
  \bibfield  {author} {\bibinfo {author} {\bibfnamefont {V.~V.}\ \bibnamefont
  {Tolmachev}},\ }\href@noop {} {\bibfield  {journal} {\bibinfo  {journal}
  {SPhD}\ }\textbf {\bibinfo {volume} {6}},\ \bibinfo {pages} {800} (\bibinfo
  {year} {1962})}\BibitemShut {NoStop}%
\bibitem [{\citenamefont {Morel}\ and\ \citenamefont
  {Anderson}(1962)}]{AndersonMorel}%
  \BibitemOpen
  \bibfield  {author} {\bibinfo {author} {\bibfnamefont {P.}~\bibnamefont
  {Morel}}\ and\ \bibinfo {author} {\bibfnamefont {P.~W.}\ \bibnamefont
  {Anderson}},\ }\href {\doibase 10.1103/PhysRev.125.1263} {\bibfield
  {journal} {\bibinfo  {journal} {Phys. Rev.}\ }\textbf {\bibinfo {volume}
  {125}},\ \bibinfo {pages} {1263} (\bibinfo {year} {1962})}\BibitemShut
  {NoStop}%
\bibitem [{\citenamefont {Nagaosa}(1999)}]{nagaosa1999quantum}%
  \BibitemOpen
  \bibfield  {author} {\bibinfo {author} {\bibfnamefont {N.}~\bibnamefont
  {Nagaosa}},\ }\href@noop {} {\emph {\bibinfo {title} {Quantum field theory in
  condensed matter physics}}}\ (\bibinfo  {publisher} {Springer Science \&
  Business Media},\ \bibinfo {year} {1999})\BibitemShut {NoStop}%
\bibitem [{\citenamefont {Cea}\ and\ \citenamefont
  {Guinea}(2021)}]{GuineaFolding}%
  \BibitemOpen
  \bibfield  {author} {\bibinfo {author} {\bibfnamefont {T.}~\bibnamefont
  {Cea}}\ and\ \bibinfo {author} {\bibfnamefont {F.}~\bibnamefont {Guinea}},\
  }\href@noop {} {\enquote {\bibinfo {title} {Coulomb interaction, phonons, and
  superconductivity in twisted bilayer graphene},}\ } (\bibinfo {year}
  {2021}),\ \Eprint {http://arxiv.org/abs/2103.01815} {arXiv:2103.01815
  [cond-mat.str-el]} \BibitemShut {NoStop}%
\bibitem [{\citenamefont {Cocemasov}\ \emph {et~al.}(2013)\citenamefont
  {Cocemasov}, \citenamefont {Nika},\ and\ \citenamefont
  {Balandin}}]{FoldedPhonons}%
  \BibitemOpen
  \bibfield  {author} {\bibinfo {author} {\bibfnamefont {A.~I.}\ \bibnamefont
  {Cocemasov}}, \bibinfo {author} {\bibfnamefont {D.~L.}\ \bibnamefont {Nika}},
  \ and\ \bibinfo {author} {\bibfnamefont {A.~A.}\ \bibnamefont {Balandin}},\
  }\href {\doibase 10.1103/PhysRevB.88.035428} {\bibfield  {journal} {\bibinfo
  {journal} {Phys. Rev. B}\ }\textbf {\bibinfo {volume} {88}},\ \bibinfo
  {pages} {035428} (\bibinfo {year} {2013})}\BibitemShut {NoStop}%
\bibitem [{\citenamefont {Halperin}\ and\ \citenamefont
  {Nelson}(1979)}]{HalperinSuperconductorResistance}%
  \BibitemOpen
  \bibfield  {author} {\bibinfo {author} {\bibfnamefont {B.~I.}\ \bibnamefont
  {Halperin}}\ and\ \bibinfo {author} {\bibfnamefont {D.~R.}\ \bibnamefont
  {Nelson}},\ }\href {\doibase 10.1007/BF00116988} {\bibfield  {journal}
  {\bibinfo  {journal} {Journal of Low Temperature Physics}\ }\textbf {\bibinfo
  {volume} {36}},\ \bibinfo {pages} {599} (\bibinfo {year} {1979})}\BibitemShut
  {NoStop}%
\bibitem [{\citenamefont {Calder\'on}\ and\ \citenamefont
  {Bascones}(2020)}]{BandStructurenoramlizations}%
  \BibitemOpen
  \bibfield  {author} {\bibinfo {author} {\bibfnamefont {M.~J.}\ \bibnamefont
  {Calder\'on}}\ and\ \bibinfo {author} {\bibfnamefont {E.}~\bibnamefont
  {Bascones}},\ }\href {\doibase 10.1103/PhysRevB.102.155149} {\bibfield
  {journal} {\bibinfo  {journal} {Phys. Rev. B}\ }\textbf {\bibinfo {volume}
  {102}},\ \bibinfo {pages} {155149} (\bibinfo {year} {2020})}\BibitemShut
  {NoStop}%
\bibitem [{\citenamefont {Goodwin}\ \emph {et~al.}(2020)\citenamefont
  {Goodwin}, \citenamefont {Vitale}, \citenamefont {Liang}, \citenamefont
  {Mostofi},\ and\ \citenamefont {Lischner}}]{HartreeBandStructure}%
  \BibitemOpen
  \bibfield  {author} {\bibinfo {author} {\bibfnamefont {Z.~A.~H.}\
  \bibnamefont {Goodwin}}, \bibinfo {author} {\bibfnamefont {V.}~\bibnamefont
  {Vitale}}, \bibinfo {author} {\bibfnamefont {X.}~\bibnamefont {Liang}},
  \bibinfo {author} {\bibfnamefont {A.~A.}\ \bibnamefont {Mostofi}}, \ and\
  \bibinfo {author} {\bibfnamefont {J.}~\bibnamefont {Lischner}},\ }\href@noop
  {} {\enquote {\bibinfo {title} {Hartree theory calculations of quasiparticle
  properties in twisted bilayer graphene},}\ } (\bibinfo {year} {2020}),\
  \Eprint {http://arxiv.org/abs/arXiv:2004.14784} {arXiv:2004.14784}
  \BibitemShut {NoStop}%
\bibitem [{\citenamefont {Xie}\ and\ \citenamefont
  {MacDonald}(2020)}]{Mcdonaldweakfieldhall}%
  \BibitemOpen
  \bibfield  {author} {\bibinfo {author} {\bibfnamefont {M.}~\bibnamefont
  {Xie}}\ and\ \bibinfo {author} {\bibfnamefont {A.~H.}\ \bibnamefont
  {MacDonald}},\ }\href@noop {} {\enquote {\bibinfo {title} {Weak-field hall
  resistivity and spin/valley flavor symmetry breaking in matbg},}\ } (\bibinfo
  {year} {2020}),\ \Eprint {http://arxiv.org/abs/arXiv:2010.07928}
  {arXiv:2010.07928} \BibitemShut {NoStop}%
\bibitem [{\citenamefont {Lewandowski}\ \emph
  {et~al.}(2021{\natexlab{b}})\citenamefont {Lewandowski}, \citenamefont
  {Nadj-Perge},\ and\ \citenamefont
  {Chowdhury}}]{FillingDependentRenormalization}%
  \BibitemOpen
  \bibfield  {author} {\bibinfo {author} {\bibfnamefont {C.}~\bibnamefont
  {Lewandowski}}, \bibinfo {author} {\bibfnamefont {S.}~\bibnamefont
  {Nadj-Perge}}, \ and\ \bibinfo {author} {\bibfnamefont {D.}~\bibnamefont
  {Chowdhury}},\ }\href@noop {} {\enquote {\bibinfo {title} {Does
  filling-dependent band renormalization aid pairing in twisted bilayer
  graphene?}}\ } (\bibinfo {year} {2021}{\natexlab{b}}),\ \Eprint
  {http://arxiv.org/abs/arXiv:2102.05661} {arXiv:2102.05661} \BibitemShut
  {NoStop}%
\bibitem [{\citenamefont {Parker}\ \emph {et~al.}(2020)\citenamefont {Parker},
  \citenamefont {Soejima}, \citenamefont {Hauschild}, \citenamefont {Zaletel},\
  and\ \citenamefont {Bultinck}}]{parker2020straininducedZalatel}%
  \BibitemOpen
  \bibfield  {author} {\bibinfo {author} {\bibfnamefont {D.~E.}\ \bibnamefont
  {Parker}}, \bibinfo {author} {\bibfnamefont {T.}~\bibnamefont {Soejima}},
  \bibinfo {author} {\bibfnamefont {J.}~\bibnamefont {Hauschild}}, \bibinfo
  {author} {\bibfnamefont {M.~P.}\ \bibnamefont {Zaletel}}, \ and\ \bibinfo
  {author} {\bibfnamefont {N.}~\bibnamefont {Bultinck}},\ }\href@noop {}
  {\enquote {\bibinfo {title} {Strain-induced quantum phase transitions in
  magic angle graphene},}\ } (\bibinfo {year} {2020}),\ \Eprint
  {http://arxiv.org/abs/2012.09885} {arXiv:2012.09885 [cond-mat.str-el]}
  \BibitemShut {NoStop}%
\bibitem [{\citenamefont {Saito}\ \emph
  {et~al.}(2020{\natexlab{b}})\citenamefont {Saito}, \citenamefont {Yang},
  \citenamefont {Ge}, \citenamefont {Liu}, \citenamefont {Watanabe},
  \citenamefont {Taniguchi}, \citenamefont {Li}, \citenamefont {Berg},\ and\
  \citenamefont {Young}}]{ISospinPomeranchuk}%
  \BibitemOpen
  \bibfield  {author} {\bibinfo {author} {\bibfnamefont {Y.}~\bibnamefont
  {Saito}}, \bibinfo {author} {\bibfnamefont {F.}~\bibnamefont {Yang}},
  \bibinfo {author} {\bibfnamefont {J.}~\bibnamefont {Ge}}, \bibinfo {author}
  {\bibfnamefont {X.}~\bibnamefont {Liu}}, \bibinfo {author} {\bibfnamefont
  {K.}~\bibnamefont {Watanabe}}, \bibinfo {author} {\bibfnamefont
  {T.}~\bibnamefont {Taniguchi}}, \bibinfo {author} {\bibfnamefont {J.~I.~A.}\
  \bibnamefont {Li}}, \bibinfo {author} {\bibfnamefont {E.}~\bibnamefont
  {Berg}}, \ and\ \bibinfo {author} {\bibfnamefont {A.~F.}\ \bibnamefont
  {Young}},\ }\href@noop {} {\enquote {\bibinfo {title} {Isospin pomeranchuk
  effect and the entropy of collective excitations in twisted bilayer
  graphene},}\ } (\bibinfo {year} {2020}{\natexlab{b}}),\ \Eprint
  {http://arxiv.org/abs/arXiv:2008.10830} {arXiv:2008.10830} \BibitemShut
  {NoStop}%
\bibitem [{\citenamefont {Rozen}\ \emph {et~al.}(2020)\citenamefont {Rozen},
  \citenamefont {Park}, \citenamefont {Zondiner}, \citenamefont {Cao},
  \citenamefont {Rodan-Legrain}, \citenamefont {Taniguchi}, \citenamefont
  {Watanabe}, \citenamefont {Oreg}, \citenamefont {Stern}, \citenamefont
  {Berg}, \citenamefont {Jarillo-Herrero},\ and\ \citenamefont
  {Ilani}}]{IlaniPomeranchuk}%
  \BibitemOpen
  \bibfield  {author} {\bibinfo {author} {\bibfnamefont {A.}~\bibnamefont
  {Rozen}}, \bibinfo {author} {\bibfnamefont {J.~M.}\ \bibnamefont {Park}},
  \bibinfo {author} {\bibfnamefont {U.}~\bibnamefont {Zondiner}}, \bibinfo
  {author} {\bibfnamefont {Y.}~\bibnamefont {Cao}}, \bibinfo {author}
  {\bibfnamefont {D.}~\bibnamefont {Rodan-Legrain}}, \bibinfo {author}
  {\bibfnamefont {T.}~\bibnamefont {Taniguchi}}, \bibinfo {author}
  {\bibfnamefont {K.}~\bibnamefont {Watanabe}}, \bibinfo {author}
  {\bibfnamefont {Y.}~\bibnamefont {Oreg}}, \bibinfo {author} {\bibfnamefont
  {A.}~\bibnamefont {Stern}}, \bibinfo {author} {\bibfnamefont
  {E.}~\bibnamefont {Berg}}, \bibinfo {author} {\bibfnamefont {P.}~\bibnamefont
  {Jarillo-Herrero}}, \ and\ \bibinfo {author} {\bibfnamefont {S.}~\bibnamefont
  {Ilani}},\ }\href@noop {} {\enquote {\bibinfo {title} {Entropic evidence for
  a pomeranchuk effect in magic angle graphene},}\ } (\bibinfo {year} {2020}),\
  \Eprint {http://arxiv.org/abs/arXiv:2009.01836} {arXiv:2009.01836}
  \BibitemShut {NoStop}%
\bibitem [{\citenamefont {Thomson}\ and\ \citenamefont
  {Alicea}(2021)}]{AliceaDisorder}%
  \BibitemOpen
  \bibfield  {author} {\bibinfo {author} {\bibfnamefont {A.}~\bibnamefont
  {Thomson}}\ and\ \bibinfo {author} {\bibfnamefont {J.}~\bibnamefont
  {Alicea}},\ }\href {\doibase 10.1103/PhysRevB.103.125138} {\bibfield
  {journal} {\bibinfo  {journal} {Phys. Rev. B}\ }\textbf {\bibinfo {volume}
  {103}},\ \bibinfo {pages} {125138} (\bibinfo {year} {2021})}\BibitemShut
  {NoStop}%
\bibitem [{\citenamefont {Chen}\ \emph {et~al.}(2019)\citenamefont {Chen},
  \citenamefont {Jiang}, \citenamefont {Wu}, \citenamefont {Lyu}, \citenamefont
  {Li}, \citenamefont {Chittari}, \citenamefont {Watanabe}, \citenamefont
  {Taniguchi}, \citenamefont {Shi}, \citenamefont {Jung}, \citenamefont
  {Zhang},\ and\ \citenamefont {Wang}}]{ABChbn}%
  \BibitemOpen
  \bibfield  {author} {\bibinfo {author} {\bibfnamefont {G.}~\bibnamefont
  {Chen}}, \bibinfo {author} {\bibfnamefont {L.}~\bibnamefont {Jiang}},
  \bibinfo {author} {\bibfnamefont {S.}~\bibnamefont {Wu}}, \bibinfo {author}
  {\bibfnamefont {B.}~\bibnamefont {Lyu}}, \bibinfo {author} {\bibfnamefont
  {H.}~\bibnamefont {Li}}, \bibinfo {author} {\bibfnamefont {B.~L.}\
  \bibnamefont {Chittari}}, \bibinfo {author} {\bibfnamefont {K.}~\bibnamefont
  {Watanabe}}, \bibinfo {author} {\bibfnamefont {T.}~\bibnamefont {Taniguchi}},
  \bibinfo {author} {\bibfnamefont {Z.}~\bibnamefont {Shi}}, \bibinfo {author}
  {\bibfnamefont {J.}~\bibnamefont {Jung}}, \bibinfo {author} {\bibfnamefont
  {Y.}~\bibnamefont {Zhang}}, \ and\ \bibinfo {author} {\bibfnamefont
  {F.}~\bibnamefont {Wang}},\ }\href {\doibase 10.1038/s41567-018-0387-2}
  {\bibfield  {journal} {\bibinfo  {journal} {Nature Physics}\ }\textbf
  {\bibinfo {volume} {15}},\ \bibinfo {pages} {237} (\bibinfo {year}
  {2019})}\BibitemShut {NoStop}%
\bibitem [{\citenamefont {He}\ \emph {et~al.}(2021)\citenamefont {He},
  \citenamefont {Li}, \citenamefont {Cai}, \citenamefont {Liu}, \citenamefont
  {Watanabe}, \citenamefont {Taniguchi}, \citenamefont {Xu},\ and\
  \citenamefont {Yankowitz}}]{TwistedDoubleBilayerYankowitz}%
  \BibitemOpen
  \bibfield  {author} {\bibinfo {author} {\bibfnamefont {M.}~\bibnamefont
  {He}}, \bibinfo {author} {\bibfnamefont {Y.}~\bibnamefont {Li}}, \bibinfo
  {author} {\bibfnamefont {J.}~\bibnamefont {Cai}}, \bibinfo {author}
  {\bibfnamefont {Y.}~\bibnamefont {Liu}}, \bibinfo {author} {\bibfnamefont
  {K.}~\bibnamefont {Watanabe}}, \bibinfo {author} {\bibfnamefont
  {T.}~\bibnamefont {Taniguchi}}, \bibinfo {author} {\bibfnamefont
  {X.}~\bibnamefont {Xu}}, \ and\ \bibinfo {author} {\bibfnamefont
  {M.}~\bibnamefont {Yankowitz}},\ }\href {\doibase 10.1038/s41567-020-1030-6}
  {\bibfield  {journal} {\bibinfo  {journal} {Nature Physics}\ }\textbf
  {\bibinfo {volume} {17}},\ \bibinfo {pages} {26} (\bibinfo {year}
  {2021})}\BibitemShut {NoStop}%
\bibitem [{\citenamefont {Shen}\ \emph {et~al.}(2020)\citenamefont {Shen},
  \citenamefont {Chu}, \citenamefont {Wu}, \citenamefont {Li}, \citenamefont
  {Wang}, \citenamefont {Zhao}, \citenamefont {Tang}, \citenamefont {Liu},
  \citenamefont {Tian}, \citenamefont {Watanabe}, \citenamefont {Taniguchi},
  \citenamefont {Yang}, \citenamefont {Meng}, \citenamefont {Shi},
  \citenamefont {Yazyev},\ and\ \citenamefont
  {Zhang}}]{TwistedDoubleBilayerShen2020}%
  \BibitemOpen
  \bibfield  {author} {\bibinfo {author} {\bibfnamefont {C.}~\bibnamefont
  {Shen}}, \bibinfo {author} {\bibfnamefont {Y.}~\bibnamefont {Chu}}, \bibinfo
  {author} {\bibfnamefont {Q.}~\bibnamefont {Wu}}, \bibinfo {author}
  {\bibfnamefont {N.}~\bibnamefont {Li}}, \bibinfo {author} {\bibfnamefont
  {S.}~\bibnamefont {Wang}}, \bibinfo {author} {\bibfnamefont {Y.}~\bibnamefont
  {Zhao}}, \bibinfo {author} {\bibfnamefont {J.}~\bibnamefont {Tang}}, \bibinfo
  {author} {\bibfnamefont {J.}~\bibnamefont {Liu}}, \bibinfo {author}
  {\bibfnamefont {J.}~\bibnamefont {Tian}}, \bibinfo {author} {\bibfnamefont
  {K.}~\bibnamefont {Watanabe}}, \bibinfo {author} {\bibfnamefont
  {T.}~\bibnamefont {Taniguchi}}, \bibinfo {author} {\bibfnamefont
  {R.}~\bibnamefont {Yang}}, \bibinfo {author} {\bibfnamefont {Z.~Y.}\
  \bibnamefont {Meng}}, \bibinfo {author} {\bibfnamefont {D.}~\bibnamefont
  {Shi}}, \bibinfo {author} {\bibfnamefont {O.~V.}\ \bibnamefont {Yazyev}}, \
  and\ \bibinfo {author} {\bibfnamefont {G.}~\bibnamefont {Zhang}},\ }\href
  {\doibase 10.1038/s41567-020-0825-9} {\bibfield  {journal} {\bibinfo
  {journal} {Nature Physics}\ }\textbf {\bibinfo {volume} {16}},\ \bibinfo
  {pages} {520} (\bibinfo {year} {2020})}\BibitemShut {NoStop}%
\bibitem [{\citenamefont {Chen}\ \emph {et~al.}(2020)\citenamefont {Chen},
  \citenamefont {Sharpe}, \citenamefont {Fox}, \citenamefont {Zhang},
  \citenamefont {Wang}, \citenamefont {Jiang}, \citenamefont {Lyu},
  \citenamefont {Li}, \citenamefont {Watanabe}, \citenamefont {Taniguchi},
  \citenamefont {Shi}, \citenamefont {Senthil}, \citenamefont
  {Goldhaber-Gordon}, \citenamefont {Zhang},\ and\ \citenamefont
  {Wang}}]{hBNgoldhaberGordon}%
  \BibitemOpen
  \bibfield  {author} {\bibinfo {author} {\bibfnamefont {G.}~\bibnamefont
  {Chen}}, \bibinfo {author} {\bibfnamefont {A.~L.}\ \bibnamefont {Sharpe}},
  \bibinfo {author} {\bibfnamefont {E.~J.}\ \bibnamefont {Fox}}, \bibinfo
  {author} {\bibfnamefont {Y.-H.}\ \bibnamefont {Zhang}}, \bibinfo {author}
  {\bibfnamefont {S.}~\bibnamefont {Wang}}, \bibinfo {author} {\bibfnamefont
  {L.}~\bibnamefont {Jiang}}, \bibinfo {author} {\bibfnamefont
  {B.}~\bibnamefont {Lyu}}, \bibinfo {author} {\bibfnamefont {H.}~\bibnamefont
  {Li}}, \bibinfo {author} {\bibfnamefont {K.}~\bibnamefont {Watanabe}},
  \bibinfo {author} {\bibfnamefont {T.}~\bibnamefont {Taniguchi}}, \bibinfo
  {author} {\bibfnamefont {Z.}~\bibnamefont {Shi}}, \bibinfo {author}
  {\bibfnamefont {T.}~\bibnamefont {Senthil}}, \bibinfo {author} {\bibfnamefont
  {D.}~\bibnamefont {Goldhaber-Gordon}}, \bibinfo {author} {\bibfnamefont
  {Y.}~\bibnamefont {Zhang}}, \ and\ \bibinfo {author} {\bibfnamefont
  {F.}~\bibnamefont {Wang}},\ }\href {\doibase 10.1038/s41586-020-2049-7}
  {\bibfield  {journal} {\bibinfo  {journal} {Nature}\ }\textbf {\bibinfo
  {volume} {579}},\ \bibinfo {pages} {56} (\bibinfo {year} {2020})}\BibitemShut
  {NoStop}%
\bibitem [{\citenamefont {Serlin}\ \emph {et~al.}(2020)\citenamefont {Serlin},
  \citenamefont {Tschirhart}, \citenamefont {Polshyn}, \citenamefont {Zhang},
  \citenamefont {Zhu}, \citenamefont {Watanabe}, \citenamefont {Taniguchi},
  \citenamefont {Balents},\ and\ \citenamefont {Young}}]{hBNyoung}%
  \BibitemOpen
  \bibfield  {author} {\bibinfo {author} {\bibfnamefont {M.}~\bibnamefont
  {Serlin}}, \bibinfo {author} {\bibfnamefont {C.~L.}\ \bibnamefont
  {Tschirhart}}, \bibinfo {author} {\bibfnamefont {H.}~\bibnamefont {Polshyn}},
  \bibinfo {author} {\bibfnamefont {Y.}~\bibnamefont {Zhang}}, \bibinfo
  {author} {\bibfnamefont {J.}~\bibnamefont {Zhu}}, \bibinfo {author}
  {\bibfnamefont {K.}~\bibnamefont {Watanabe}}, \bibinfo {author}
  {\bibfnamefont {T.}~\bibnamefont {Taniguchi}}, \bibinfo {author}
  {\bibfnamefont {L.}~\bibnamefont {Balents}}, \ and\ \bibinfo {author}
  {\bibfnamefont {A.~F.}\ \bibnamefont {Young}},\ }\href {\doibase
  10.1126/science.aay5533} {\bibfield  {journal} {\bibinfo  {journal}
  {Science}\ }\textbf {\bibinfo {volume} {367}},\ \bibinfo {pages} {900}
  (\bibinfo {year} {2020})},\ \Eprint
  {http://arxiv.org/abs/https://science.sciencemag.org/content/367/6480/900.full.pdf}
  {https://science.sciencemag.org/content/367/6480/900.full.pdf} \BibitemShut
  {NoStop}%
\bibitem [{\citenamefont {Park}\ \emph {et~al.}(2020)\citenamefont {Park},
  \citenamefont {Cao}, \citenamefont {Watanabe}, \citenamefont {Taniguchi},\
  and\ \citenamefont {Jarillo-Herrero}}]{TrilayerPablo}%
  \BibitemOpen
  \bibfield  {author} {\bibinfo {author} {\bibfnamefont {J.~M.}\ \bibnamefont
  {Park}}, \bibinfo {author} {\bibfnamefont {Y.}~\bibnamefont {Cao}}, \bibinfo
  {author} {\bibfnamefont {K.}~\bibnamefont {Watanabe}}, \bibinfo {author}
  {\bibfnamefont {T.}~\bibnamefont {Taniguchi}}, \ and\ \bibinfo {author}
  {\bibfnamefont {P.}~\bibnamefont {Jarillo-Herrero}},\ }\href@noop {}
  {\enquote {\bibinfo {title} {Tunable phase boundaries and ultra-strong
  coupling superconductivity in mirror symmetric magic-angle trilayer
  graphene},}\ } (\bibinfo {year} {2020}),\ \Eprint
  {http://arxiv.org/abs/arXiv:2012.01434} {arXiv:2012.01434} \BibitemShut
  {NoStop}%
\bibitem [{\citenamefont {Hao}\ \emph {et~al.}(2020)\citenamefont {Hao},
  \citenamefont {Zimmerman}, \citenamefont {Ledwith}, \citenamefont {Khalaf},
  \citenamefont {Najafabadi}, \citenamefont {Watanabe}, \citenamefont
  {Taniguchi}, \citenamefont {Vishwanath},\ and\ \citenamefont
  {Kim}}]{TrilayerKim}%
  \BibitemOpen
  \bibfield  {author} {\bibinfo {author} {\bibfnamefont {Z.}~\bibnamefont
  {Hao}}, \bibinfo {author} {\bibfnamefont {A.~M.}\ \bibnamefont {Zimmerman}},
  \bibinfo {author} {\bibfnamefont {P.}~\bibnamefont {Ledwith}}, \bibinfo
  {author} {\bibfnamefont {E.}~\bibnamefont {Khalaf}}, \bibinfo {author}
  {\bibfnamefont {D.~H.}\ \bibnamefont {Najafabadi}}, \bibinfo {author}
  {\bibfnamefont {K.}~\bibnamefont {Watanabe}}, \bibinfo {author}
  {\bibfnamefont {T.}~\bibnamefont {Taniguchi}}, \bibinfo {author}
  {\bibfnamefont {A.}~\bibnamefont {Vishwanath}}, \ and\ \bibinfo {author}
  {\bibfnamefont {P.}~\bibnamefont {Kim}},\ }\href@noop {} {\enquote {\bibinfo
  {title} {Electric field tunable unconventional superconductivity in
  alternating twist magic-angle trilayer graphene},}\ } (\bibinfo {year}
  {2020}),\ \Eprint {http://arxiv.org/abs/arXiv:2012.02773} {arXiv:2012.02773}
  \BibitemShut {NoStop}%
\bibitem [{\citenamefont {Allen}\ and\ \citenamefont
  {Dynes}(1975)}]{AllenDynesLinearinLambdaTc}%
  \BibitemOpen
  \bibfield  {author} {\bibinfo {author} {\bibfnamefont {P.~B.}\ \bibnamefont
  {Allen}}\ and\ \bibinfo {author} {\bibfnamefont {R.~C.}\ \bibnamefont
  {Dynes}},\ }\href {\doibase 10.1103/PhysRevB.12.905} {\bibfield  {journal}
  {\bibinfo  {journal} {Phys. Rev. B}\ }\textbf {\bibinfo {volume} {12}},\
  \bibinfo {pages} {905} (\bibinfo {year} {1975})}\BibitemShut {NoStop}%
\bibitem [{\citenamefont {Kopnin}\ and\ \citenamefont
  {Sonin}(2008)}]{GrapheneBCS}%
  \BibitemOpen
  \bibfield  {author} {\bibinfo {author} {\bibfnamefont {N.~B.}\ \bibnamefont
  {Kopnin}}\ and\ \bibinfo {author} {\bibfnamefont {E.~B.}\ \bibnamefont
  {Sonin}},\ }\href {\doibase 10.1103/PhysRevLett.100.246808} {\bibfield
  {journal} {\bibinfo  {journal} {Phys. Rev. Lett.}\ }\textbf {\bibinfo
  {volume} {100}},\ \bibinfo {pages} {246808} (\bibinfo {year}
  {2008})}\BibitemShut {NoStop}%
\bibitem [{\citenamefont {Abrikosov}(2017)}]{abrikosov2017fundamentals}%
  \BibitemOpen
  \bibfield  {author} {\bibinfo {author} {\bibfnamefont {A.}~\bibnamefont
  {Abrikosov}},\ }\href {https://books.google.co.il/books?id=tTo2DwAAQBAJ}
  {\emph {\bibinfo {title} {Fundamentals of the Theory of Metals}}}\ (\bibinfo
  {publisher} {Dover Publications},\ \bibinfo {year} {2017})\BibitemShut
  {NoStop}%
\bibitem [{\citenamefont {Aslamasov}\ and\ \citenamefont
  {Larkin}(1968)}]{ASLAMASOV1968238}%
  \BibitemOpen
  \bibfield  {author} {\bibinfo {author} {\bibfnamefont {L.}~\bibnamefont
  {Aslamasov}}\ and\ \bibinfo {author} {\bibfnamefont {A.}~\bibnamefont
  {Larkin}},\ }\href {\doibase https://doi.org/10.1016/0375-9601(68)90623-3}
  {\bibfield  {journal} {\bibinfo  {journal} {Physics Letters A}\ }\textbf
  {\bibinfo {volume} {26}},\ \bibinfo {pages} {238 } (\bibinfo {year}
  {1968})}\BibitemShut {NoStop}%
\bibitem [{\citenamefont {Kim}\ \emph {et~al.}(2018)\citenamefont {Kim},
  \citenamefont {Leconte}, \citenamefont {Chittari}, \citenamefont {Watanabe},
  \citenamefont {Taniguchi}, \citenamefont {MacDonald}, \citenamefont {Jung},\
  and\ \citenamefont {Jung}}]{hBNgapsize}%
  \BibitemOpen
  \bibfield  {author} {\bibinfo {author} {\bibfnamefont {H.}~\bibnamefont
  {Kim}}, \bibinfo {author} {\bibfnamefont {N.}~\bibnamefont {Leconte}},
  \bibinfo {author} {\bibfnamefont {B.~L.}\ \bibnamefont {Chittari}}, \bibinfo
  {author} {\bibfnamefont {K.}~\bibnamefont {Watanabe}}, \bibinfo {author}
  {\bibfnamefont {T.}~\bibnamefont {Taniguchi}}, \bibinfo {author}
  {\bibfnamefont {A.~H.}\ \bibnamefont {MacDonald}}, \bibinfo {author}
  {\bibfnamefont {J.}~\bibnamefont {Jung}}, \ and\ \bibinfo {author}
  {\bibfnamefont {S.}~\bibnamefont {Jung}},\ }\href {\doibase
  10.1021/acs.nanolett.8b03423} {\bibfield  {journal} {\bibinfo  {journal}
  {Nano Letters}\ }\textbf {\bibinfo {volume} {18}},\ \bibinfo {pages} {7732}
  (\bibinfo {year} {2018})},\ \bibinfo {note} {pMID: 30457338},\ \Eprint
  {http://arxiv.org/abs/https://doi.org/10.1021/acs.nanolett.8b03423}
  {https://doi.org/10.1021/acs.nanolett.8b03423} \BibitemShut {NoStop}%
\bibitem [{\citenamefont {Akera}\ \emph {et~al.}(1991)\citenamefont {Akera},
  \citenamefont {MacDonald}, \citenamefont {Girvin},\ and\ \citenamefont
  {Norman}}]{VortexLattice}%
  \BibitemOpen
  \bibfield  {author} {\bibinfo {author} {\bibfnamefont {H.}~\bibnamefont
  {Akera}}, \bibinfo {author} {\bibfnamefont {A.~H.}\ \bibnamefont
  {MacDonald}}, \bibinfo {author} {\bibfnamefont {S.~M.}\ \bibnamefont
  {Girvin}}, \ and\ \bibinfo {author} {\bibfnamefont {M.~R.}\ \bibnamefont
  {Norman}},\ }\href {\doibase 10.1103/PhysRevLett.67.2375} {\bibfield
  {journal} {\bibinfo  {journal} {Phys. Rev. Lett.}\ }\textbf {\bibinfo
  {volume} {67}},\ \bibinfo {pages} {2375} (\bibinfo {year}
  {1991})}\BibitemShut {NoStop}%
\bibitem [{\citenamefont {Maki}(1969)}]{maki1969gapless}%
  \BibitemOpen
  \bibfield  {author} {\bibinfo {author} {\bibfnamefont {K.}~\bibnamefont
  {Maki}},\ }\href@noop {} {\bibfield  {journal} {\bibinfo  {journal}
  {Superconductivity: Part}\ }\textbf {\bibinfo {volume} {2}} (\bibinfo {year}
  {1969})}\BibitemShut {NoStop}%
\bibitem [{\citenamefont {Nomura}\ and\ \citenamefont
  {MacDonald}(2006)}]{MacdonaldQHFM}%
  \BibitemOpen
  \bibfield  {author} {\bibinfo {author} {\bibfnamefont {K.}~\bibnamefont
  {Nomura}}\ and\ \bibinfo {author} {\bibfnamefont {A.~H.}\ \bibnamefont
  {MacDonald}},\ }\href {\doibase 10.1103/PhysRevLett.96.256602} {\bibfield
  {journal} {\bibinfo  {journal} {Phys. Rev. Lett.}\ }\textbf {\bibinfo
  {volume} {96}},\ \bibinfo {pages} {256602} (\bibinfo {year}
  {2006})}\BibitemShut {NoStop}%
\bibitem [{\citenamefont {Kharitonov}(2012)}]{KharitonovGraphene}%
  \BibitemOpen
  \bibfield  {author} {\bibinfo {author} {\bibfnamefont {M.}~\bibnamefont
  {Kharitonov}},\ }\href {\doibase 10.1103/PhysRevB.85.155439} {\bibfield
  {journal} {\bibinfo  {journal} {Phys. Rev. B}\ }\textbf {\bibinfo {volume}
  {85}},\ \bibinfo {pages} {155439} (\bibinfo {year} {2012})}\BibitemShut
  {NoStop}%
\bibitem [{\citenamefont {Bultinck}\ \emph
  {et~al.}(2020{\natexlab{b}})\citenamefont {Bultinck}, \citenamefont
  {Chatterjee},\ and\ \citenamefont {Zaletel}}]{AHEprlBultnickZalatel}%
  \BibitemOpen
  \bibfield  {author} {\bibinfo {author} {\bibfnamefont {N.}~\bibnamefont
  {Bultinck}}, \bibinfo {author} {\bibfnamefont {S.}~\bibnamefont
  {Chatterjee}}, \ and\ \bibinfo {author} {\bibfnamefont {M.~P.}\ \bibnamefont
  {Zaletel}},\ }\href {\doibase 10.1103/PhysRevLett.124.166601} {\bibfield
  {journal} {\bibinfo  {journal} {Phys. Rev. Lett.}\ }\textbf {\bibinfo
  {volume} {124}},\ \bibinfo {pages} {166601} (\bibinfo {year}
  {2020}{\natexlab{b}})}\BibitemShut {NoStop}%
\end{thebibliography}%

\begin{widetext}

\global\long\def\thesection{S.\Alph{section}}%
 \setcounter{figure}{0} 
\global\long\def\thefigure{S\arabic{figure}}%
 \setcounter{equation}{0} 
\global\long\def\theequation{S\arabic{equation}}%
\section*{Supplemental Material for ``Theory of correlated insulators and superconductivity in twisted bilayer graphene''}

\setcounter{section}{0} \renewcommand{\thesection}{S.\arabic{section}} \setcounter{figure}{0} \renewcommand{\thefigure}{S\arabic{figure}} \setcounter{equation}{0} \renewcommand{\theequation}{S\arabic{equation}}

\section {The proposed model and mean-field variational approach}\label{app:modelandvariation}

\subsection {Model description}\label{app:modeldescription}

In this work we explore a model comprised of eight flat bands with
valley ($K/K'$), spin, and sublattice ($A/B$) degrees of freedom,
labeled by Pauli matrices $\tau_{i}$, $s_{i}$, and $\sigma_{i}$,
respectively. This choice of basis is motivated by the MATBG sublattice-polarized
basis which adiabatically connects the fully-polarized bands in the
chiral limit to the realistic model \cite{BultnickKhalaf2020}. These
bands have a valley-dependent Chern number, $C=\sigma_{z}\tau_{z}$,
which can be understood from the two chiral Dirac points in each valley.
Using the 8-spinor 
\begin{equation}\label{eq:Psidefine}
\Psi=\left(c_{K\uparrow A},c_{K\uparrow B},c_{K'\uparrow  A},c_{K'\uparrow B},c_{K\downarrow A},c_{K\downarrow B},c_{K'\downarrow A},c_{K'\downarrow B}\right)^{{\rm T}},
\end{equation}
We write the Hamiltonian as
\begin{equation}
H=\sum_{\mathbf{k}}\Psi_{\mathbf{k}}^{\dagger}h_{0}\left(\mathbf{k}\right)\Psi_{\mathbf{k}}+H_{{\rm int}},\label{eq:ModelHamiltonianSupp}
\end{equation}
where the single-particle part is of the form
\begin{equation}
h_{0}\left(\mathbf{k}\right)=f_{x}\left(\mathbf{k}\right)\sigma_{x}+f_{y}\left(\mathbf{k}\right)\sigma_{y}\tau_{z} + f_{\rm p-h}\left(\mathbf{k}\right).
\end{equation}
In this work, we use
\begin{equation}
    f_{x}\left(\mathbf{k}\right)=\frac{1}{2M_{\rm eff}}\left[k_x^2 - k_y ^2- \tilde{k}^2\right],
\end{equation}
\begin{equation}
    f_{y}\left(\mathbf{k}\right) =\frac{1}{2M_{\rm eff}}\left[ 2 k_x k_y\right],
\end{equation}
and work in units where $2M_{\rm eff}=1$ for convenience.
This produces an energy spectrum with two Dirac cones of the same chirality at momenta $\left(k_x ,k_y\right)= \left(\pm\tilde{k} ,0\right)$, allowing us to emulate the structure of the isolated flat bands in magic angle twisted bilayer graphene (MATBG), including the same topological structure of Chern bands once a perturbation $\propto \sigma_z$ is introduced, opening a band gap.
The limits of our effective Brillouin are $k_x\in\left[ -0.56, 0.56\right]$ and $k_x\in\left[ -0.26, 0.26\right]$. We also use $\tilde{k}=0.5$.

The last term in the single-particle part of the Hamiltonian introduces particle-hole asymmetry into the model.
In the case of a single Dirac cone with a spectrum $\propto \pm \left|{\mathbf{k}}\right|$, the simplest way to include particle hole asymmetry is to add a quadratic term $\propto \left|{\mathbf{k}}\right|^2$ which induces curvature in the Dirac dispersion. Since we wish to introduce such a curvature near \textit{both} the Dirac points, we choose to use 
\begin{equation}
    f_{\rm p-h}\left(\mathbf{k}\right) =\frac{1}{2m_{\rm p-h}}\sqrt{\left(k_x-\tilde{k}\right)^2+k_y ^2}\sqrt{\left(k_x+\tilde{k}\right)^2+k_y ^2},\label{eq:phterm}
\end{equation}
which indeed has the desired effect on the electronic density of states, see Fig.~\ref{fig:particleholesupplemental}.
Eq.~\eqref{eq:phterm} has spurious singularities at the Dirac points. We have checked that our results do not change qualitatively if we use different forms of the particle-hole symmetry breaking term, as long as the overall DOS remains similar.
Since experimental evidence suggest the conduction band is flatter as compared to the valence band \cite{YankowitzTuningMATBG,DiracRevivals,Mcdonaldweakfieldhall}, we choose the sign of $f_{\rm p-h}$ to be negative throughout the mBZ. Throughout our calculations $m_{\rm p-h}=-2.4$ is used

We stress that the specific details of $h_0 \left(\mathbf{k}\right)$ are not important within our phenomenological model, where due to the mean-field nature of our treatment and the locality of the interactions, only the DOS is important in determining the phase diagram.

We define the combined bandwidth of the conduction and valence bands of the single-particle Hamiltonian $h_0$ as $W$. This energy scale is our reference point to which we compare interaction energies, transition temperatures and gap sizes.

The interaction part of the Hamiltonian can be written out as a sum of four contributions,
\begin{align}
    H_{{\rm int}}&=\frac{U_{{\rm C}}}{2\Omega}\sum_{\mathbf{k,k',q}}\Psi_{\mathbf{k+q}}^{\dagger}\Psi_{\mathbf{k}}\Psi_{\mathbf{k'-q}}^{\dagger}\Psi_{\mathbf{k'}}\nonumber\\
    &+\frac{U_{\delta}}{2\Omega}\sum_{\mathbf{k,k',q}}\left[\Psi_{\mathbf{k+q}}^{\dagger}\sigma_{x}\tau_{z}\Psi_{\mathbf{k}}\Psi_{\mathbf{k'-q}}^{\dagger}\sigma_{x}\tau_{z}\Psi_{\mathbf{k'}}+\Psi_{\mathbf{k+q}}^{\dagger}\sigma_{y}\Psi_{\mathbf{k}}\Psi_{\mathbf{k'-q}}^{\dagger}\sigma_{y}\Psi_{\mathbf{k'}}\right]\nonumber\\
    &+\frac{-\left|g_1\right|}{2\Omega}\sum_{\mathbf{k,k',q}}\left[\Psi_{\mathbf{k+q}}^{\dagger}\sigma_{y}\tau_{z}\Psi_{\mathbf{k}}\Psi_{\mathbf{k'-q}}^{\dagger}\sigma_{y}\tau_{z}\Psi_{\mathbf{k'}}+\Psi_{\mathbf{k+q}}^{\dagger}\sigma_{x}\Psi_{\mathbf{k}}\Psi_{\mathbf{k'-q}}^{\dagger}\sigma_{x}\Psi_{\mathbf{k'}}\right]\nonumber\\
    &+\frac{-\left|g_2\right|}{2\Omega}\sum_{\mathbf{k,k',q}}\left[\Psi_{\mathbf{k+q}}^{\dagger}\sigma_{x}\tau_{x}\Psi_{\mathbf{k}}\Psi_{\mathbf{k'-q}}^{\dagger}\sigma_{x}\tau_{x}\Psi_{\mathbf{k'}}+\Psi_{\mathbf{k+q}}^{\dagger}\sigma_{x}\tau_{y}\Psi_{\mathbf{k}}\Psi_{\mathbf{k'-q}}^{\dagger}\sigma_{x}\tau_{y}\Psi_{\mathbf{k'}}\right].\label{eq:HintSupp}
\end{align}

Let us briefly discuss the origin of each of the terms in $H_{\rm int}$. The first and most dominant $U_{\rm C}$ term, is the ``symmetric'' or structure-less density-density interaction due to the short range part of the screened Coulomb interaction.
The second sub-dominant term $U_\delta$ has unusual structure in valley-sublattice space. It represents the part of the flat-band-projected Coulomb interactions which anti-commute with the chirality operator $\sigma_z$. As discussed in Ref.~\cite{BultnickKhalaf2020},
its structure is constrained by the symmetries of MATBG.

The next two terms, proportional to $g_1$ and $g_2$, originate in electron-phonon interactions. The relevant phonons here are the \textit{optical} in-plane phonons of monolayer graphene. Due to their high frequency, the lead to effective instantaneous electron-electron interactions. 
The structure of the interactions in valley-sublattice space is inherited from the structure of the electron-phonon couplings of the relevant modes. More concretely, $g_1$ terms are due to coupling of phonons with zero momentum, whereas $g_2$ originates in coupling to phonon branches located near the $K$ and $K'$ points of the monolayer graphene Brillouin zone. Thus, the $g_2$ interactions contain inter-valley scattering ($\tau_x/\tau_y$ elements), whereas $g_1$ interactions are strictly intra-valley in nature.

\subsection {Variational mean-field approach}\label{app:variationalapproach}
We begin by making an ansatz for a mean-field quadratic Hamiltonian
\begin{equation}
H_{{\rm MF}}=\sum_{\mathbf{k}}\Psi_{\mathbf{k}}^{\dagger}h_{{\rm MF}}\left(\mathbf{k}\right)\Psi_{\mathbf{k}}=\sum_{\mathbf{k},\lambda}\xi_{\mathbf{k},\lambda}\psi_{\mathbf{k},\lambda}^{\dagger}\psi_{\mathbf{k},\lambda},
\end{equation}
where we have written $H_{{\rm MF}}$ in a new diagonal basis on the
right hand side, with $\psi_{\mathbf{k},\lambda}$ a fermionic annihilation
operator with momentum $\mathbf{k}$, and $\lambda$ labels the eight
eigenstates per $\mathbf{k}$. Expectation values calculated within
the distribution generated by $H_{{\rm MF}}$ follow
\begin{equation}
\left\langle \psi_{\mathbf{k},\lambda}^{\dagger}\psi_{\mathbf{k'},\lambda'}\right\rangle _{{\rm MF}}=\delta_{\mathbf{k},\mathbf{k'}}\delta_{\lambda,\lambda'}f\left(\beta\xi_{\mathbf{k},\lambda}\right),\label{eq:FDexpectation values}
\end{equation}
with $f\left(x\right)=\left(1+e^{x}\right)^{-1}$ the Fermi-Dirac
distribution, and $\beta$ is the inverse temperature.

Our aim is to find the mean-field ansatz which minimizes the grand potential
\begin{equation}
\Phi=\left\langle H-\mu N\right\rangle _{{\rm MF}}-T\sum_{\mathbf{k},\lambda}\ln\left(1+e^{-\beta\left|\xi_{\mathbf{k},\lambda}\right|}\right),\label{eq:GibbsFreeEnergy}
\end{equation}
with $N=\sum_{\mathbf{k}}\Psi_{\mathbf{k}}^{\dagger}\Psi_{\mathbf{k}}=\sum_{\mathbf{k},\lambda}\psi_{\mathbf{k},\lambda}^{\dagger}\psi_{\mathbf{k},\lambda}$
the particle number operator, and $\mu$ the chemical potential.

The only non-quadratic part in the evaluation of Eq. (\ref{eq:GibbsFreeEnergy})
is $\left\langle H_{{\rm int}}\right\rangle _{{\rm MF}}$, for which
we can employ Wick's theorem since $H_{{\rm MF}}$ is quadratic in
fermion operators. Beginning with the dominant ``structure-less''
term, $H_{{\rm int},{\rm C}}=\frac{U_{{\rm C}}}{2\Omega}\sum_{\mathbf{k,k',q}}\Psi_{\mathbf{k+q}}^{\dagger}\Psi_{\mathbf{k}}\Psi_{\mathbf{k'-q}}^{\dagger}\Psi_{\mathbf{k'}}$, we find (we omit the ${\rm MF}$ subscript from $\left\langle \cdot\right\rangle _{{\rm MF}}$
for simplicity henceforth) 

\begin{align}
\frac{\left\langle H_{{\rm int},{\rm C}}\right\rangle }{\Omega} & =\frac{U_{{\rm C}}}{2\Omega^{2}}\sum_{\mathbf{kk'q}}\sum_{\tau's'\sigma'}\sum_{\tau s\sigma}\left\langle c_{\tau s\sigma}^{\dagger}\left(\mathbf{k+q}\right)c_{\tau s\sigma}\left(\mathbf{k}\right)c_{\tau's'\sigma'}^{\dagger}\left(\mathbf{k'-q}\right)c_{\tau's'\sigma'}\left(\mathbf{k'}\right)\right\rangle \nonumber \\
 & =\frac{U_{{\rm C}}}{2}\left[\sum_{\mathbf{k}\tau s\sigma}\left\langle c_{\tau s\sigma}^{\dagger}\left(\mathbf{k}\right)c_{\tau s\sigma}\left(\mathbf{k}\right)\right\rangle \right]^{2}\nonumber \\
 & +\frac{U_{{\rm C}}}{2}\sum_{\tau s\sigma}\left[\frac{1}{\Omega}\sum_{\mathbf{k}}\left\langle c_{\tau s\sigma}^{\dagger}\left(\mathbf{k}\right)c_{\tau s\sigma}\left(\mathbf{k}\right)\right\rangle \right]\left[\frac{1}{\Omega}\sum_{\mathbf{k'}}\left\langle 1-c_{\tau s\sigma}^{\dagger}\left(\mathbf{k'}\right)c_{\tau s\sigma}\left(\mathbf{k'}\right)\right\rangle \right]\nonumber \\
 & -\frac{U_{{\rm C}}}{2}\sum_{\tau s\sigma}\left|\frac{1}{\Omega}\sum_{\mathbf{k}}\left\langle c_{\tau s\sigma}^{\dagger}\left(\mathbf{k'}\right)c_{\tau s\bar{\sigma}}\left(\mathbf{k'}\right)\right\rangle \right|^{2}\nonumber \\
 & -\frac{U_{{\rm C}}}{2}\sum_{\tau s\sigma s'}\left|\frac{1}{\Omega}\sum_{\mathbf{k}}\left\langle c_{\tau s\sigma}^{\dagger}\left(\mathbf{k}\right)c_{\bar{\tau}s'\bar{\sigma}}\left(\mathbf{k}\right)\right\rangle \right|^{2}.\label{eq:UcExplicit}
\end{align}
The second to last term vanishes due to the chirality of the single-particle
Hamiltonian, such that summation over the entire Brillouin zone of
the term $\left\langle c_{\tau sA}^{\dagger}\left(\mathbf{k}\right)c_{\tau sB}\left(\mathbf{k}\right)\right\rangle$
is zero. The first two terms in Eq. (\ref{eq:UcExplicit})
may lead to generalized Stoner instabilities, given $U_{{\rm C}}$
is sufficiently strong as compared to the bandwidth. 

The last term
indicates that this interaction terms also favors the formation of
any inter-valley coherent (IVC) order. This order may be understood as a sort of Stoner instability as well,
where the valley $U_{\rm v}\left(1\right)$ symmetry is spontaneously broken. We have also implicitly assumed the
absence of any IVC order between two bands with opposite Chern numbers $C=\sigma_z \tau_z$, as such orders are prohibited from being uniform in $\mathbf{k}$. Instead, one expects to find an Abrikosov-like vortex lattice structure of this order parameter (reminiscent of intra-Landau-level superconductivity \cite{VortexLattice}), which comes with great kinetic energy cost~ \cite{AHEprlBultnickZalatel}. In this work we assume that this energy cost suppresses these kinds of IVC order in the mean-field state.

We follow the same procedure for the secondary interaction terms,
and we find ($H_{{\rm int},\delta}$, $H_{{\rm int,ph},1}$, and $H_{{\rm int,ph},2}$ are the parts of $H_{\rm int}$ proportional to $U_\delta$, $g_1$, and $g_2$, respectively)
\begin{align}
\frac{\left\langle H_{{\rm int},\delta}\right\rangle }{\Omega} & =U_{\delta}\sum_{\tau s\sigma}\left[\frac{1}{\Omega}\sum_{\mathbf{k}}\left\langle c_{\tau s\sigma}^{\dagger}\left(\mathbf{k}\right)c_{\tau s\sigma}\left(\mathbf{k}\right)\right\rangle \right]\left[\frac{1}{\Omega}\sum_{\mathbf{k'}}\left\langle 1-c_{\tau s\bar{\sigma}}^{\dagger}\left(\mathbf{k'}\right)c_{\tau s\bar{\sigma}}\left(\mathbf{k'}\right)\right\rangle \right]\nonumber \\
 & +U_{\delta}\sum_{\tau s\sigma}\left[\frac{1}{\Omega}\sum_{\mathbf{k}}\left\langle c_{\tau s\sigma}^{\dagger}\left(\mathbf{k}\right)c_{\bar{\tau}\bar{s}\bar{\sigma}}\left(\mathbf{k}\right)\right\rangle \right]\left[\frac{1}{\Omega}\sum_{\mathbf{k'}}\left\langle c_{\bar{\tau}\bar{s}\sigma}^{\dagger}\left(\mathbf{k'}\right)c_{\tau s\bar{\sigma}}\left(\mathbf{k'}\right)\right\rangle \right]\nonumber \\
 & +U_{\delta}\sum_{\tau s\sigma}\left[\frac{1}{\Omega}\sum_{\mathbf{k}}\left\langle c_{\tau s\sigma}^{\dagger}\left(\mathbf{k}\right)c_{\bar{\tau}s\bar{\sigma}}\left(\mathbf{k}\right)\right\rangle \right]\left[\frac{1}{\Omega}\sum_{\mathbf{k'}}\left\langle c_{\bar{\tau}s\sigma}^{\dagger}\left(\mathbf{k'}\right)c_{\tau s\bar{\sigma}}\left(\mathbf{k'}\right)\right\rangle \right],\label{eq:Hintdeltaexp}
\end{align}
\begin{align}
\frac{\left\langle H_{{\rm int},{\rm ph},1}\right\rangle }{\Omega} & =-\left|g_1\right|\sum_{\tau s\sigma}\left[\frac{1}{\Omega}\sum_{\mathbf{k}}\left\langle c_{\tau s\sigma}^{\dagger}\left(\mathbf{k}\right)c_{\tau s\sigma}\left(\mathbf{k}\right)\right\rangle \right]\left[\frac{1}{\Omega}\sum_{\mathbf{k'}}\left\langle 1-c_{\tau s\bar{\sigma}}^{\dagger}\left(\mathbf{k'}\right)c_{\tau s\bar{\sigma}}\left(\mathbf{k'}\right)\right\rangle \right]\nonumber \\
 & +\left|g_1\right|\sum_{\tau s\sigma}\left[\frac{1}{\Omega}\sum_{\mathbf{k}}\left\langle c_{\tau s\sigma}^{\dagger}\left(\mathbf{k}\right)c_{\bar{\tau}\bar{s}\bar{\sigma}}\left(\mathbf{k}\right)\right\rangle \right]\left[\frac{1}{\Omega}\sum_{\mathbf{k'}}\left\langle c_{\bar{\tau}\bar{s}\sigma}^{\dagger}\left(\mathbf{k'}\right)c_{\tau s\bar{\sigma}}\left(\mathbf{k'}\right)\right\rangle \right]\nonumber \\
 & +\left|g_1\right|\sum_{\tau s\sigma}\left[\frac{1}{\Omega}\sum_{\mathbf{k}}\left\langle c_{\tau s\sigma}^{\dagger}\left(\mathbf{k}\right)c_{\bar{\tau}s\bar{\sigma}}\left(\mathbf{k}\right)\right\rangle \right]\left[\frac{1}{\Omega}\sum_{\mathbf{k'}}\left\langle c_{\bar{\tau}s\sigma}^{\dagger}\left(\mathbf{k'}\right)c_{\tau s\bar{\sigma}}\left(\mathbf{k'}\right)\right\rangle \right],\label{eq:Hintph1exp}
\end{align}
\begin{align}
\frac{\left\langle H_{{\rm int},{\rm ph},2}\right\rangle }{\Omega} & =-\left|g_2\right|\sum_{\tau s\sigma}\left[\frac{1}{\Omega}\sum_{\mathbf{k}}\left\langle c_{\tau s\sigma}^{\dagger}\left(\mathbf{k}\right)c_{\tau s\sigma}\left(\mathbf{k}\right)\right\rangle \right]\left[\frac{1}{\Omega}\sum_{\mathbf{k'}}\left\langle 1-c_{\bar{\tau}s\bar{\sigma}}^{\dagger}\left(\mathbf{k'}\right)c_{\bar{\tau}s\bar{\sigma}}\left(\mathbf{k'}\right)\right\rangle \right]\nonumber \\
 & -2\left|g_2\right|\left|\sum_{s\sigma}\frac{1}{\Omega}\sum_{\mathbf{k}}\left\langle c_{Ks\sigma}^{\dagger}\left(\mathbf{k}\right)c_{K's\bar{\sigma}}\left(\mathbf{k}\right)\right\rangle \right|^{2}.\label{eq:Hintph2exp}
\end{align}

From the first line of Eq. (\ref{eq:Hintdeltaexp}) and the first
line of Eq. (\ref{eq:Hintph1exp}) we can understand that $U_{\delta}$
suppresses intra-flavor sublattice symmetry breaking ($\sigma_{z}$
terms), whereas $g_1$ promotes such ordered states. This can be understood from examining the sign of the interaction between the mean-field densities of the same spin-valley flavor on opposite sub-lattices. 

The second line in both equations addresses opposite-spin IVC orders.
It appears both $U_{\delta}$ and $g_1$ favor IVCs which
have $\left\langle c_{\tau sA}^{\dagger}c_{\bar{\tau}s'B}\right\rangle =-\left\langle c_{\tau sB}^{\dagger}c_{\bar{\tau}s'A}\right\rangle $,
leading to an order parameter proportional to $\sigma_{y}$. As we see from the last lines in (\ref{eq:Hintdeltaexp})--(\ref{eq:Hintph1exp}),
this same statement is true for same-spin IVC. 

However, this same-spin IVC order is suppressed due to
the last line of Eq. \eqref{eq:Hintph2exp}, which favors $\sigma_{x}$
alignment within each IVC sector, as well as alignment between the two sectors. 
In order to avoid this competition, which exists only when one considers same-spin IVC order, we focus our attention on opposite-spin IVC orders.

Moreover, the first line of Eq. \eqref{eq:Hintph2exp} explicitly favors inter-valley antiferromagnetism whenever the spin $SU\left(2\right)$ is broken, further bolstering our conclusions regarding the type of IVC one expects this model to favor.

To conclude, analytical examination of the mean-field energetics motivates us to write the variational ansatz Hamiltonian with the following form,
\begin{equation}
H_{{\rm MF}}=\sum_{\mathbf{k}}\Psi_{\mathbf{k}}^{\dagger}\left[h_{0}\left(\mathbf{k}\right)+\begin{pmatrix}\mu_{1}+m_{1} &  &  &  &  &  &  & \Delta_{{\rm ivc}}^+\\
 & \mu_{1}-m_{1} &  &  &  &  & -\Delta_{{\rm ivc}}^+\\
 &  & \mu_{2}+m_{2} &  &  & \Delta_{{\rm ivc}}^-\\
 &  &  & \mu_{2}-m_{2} & -\Delta_{{\rm ivc}}^-\\
 &  &  & -\left(\Delta_{{\rm ivc}}^-\right)^{*} & \mu_{3}+m_{3}\\
 &  & \left(\Delta_{{\rm ivc}}^-\right)^{*} &  &  & \mu_{3}-m_{3}\\
 & -\left(\Delta_{{\rm ivc}}^+\right)^{*} &  &  &  &  & \mu_{4}+m_{4}\\
\left(\Delta_{{\rm ivc}}^+\right)^{*} &  &  &  &  &  &  & \mu_{4}-m_{4}
\end{pmatrix}\right]\Psi_{\mathbf{k}},\label{eq:MFhamiltonianansztz}
\end{equation}
where $\mu_{i}$ account for flavor-symmetry breaking, $m_{i}$ correspond
to intra-flavor Chern gaps, and $\Delta_{{\rm ivc}}^\pm$ are the
relevant inter-valley coherent terms. 
We remind the reader that in this matrix the blocks numbered $\left[1,2,3,4\right]$ correspond to the flavor labels $\left[K\uparrow,K'\uparrow,K\downarrow,K'\downarrow\right]$, respectively, as can be understood from the explicit from of $\Psi_{\mathbf{k}}$ in Eq.~\eqref{eq:Psidefine}.
Upon diagonalization of $H_{{\rm MF}}$,
$\Phi$ may be calculated and minimized as a function of the variational parameters $\left\{ \mu_{i},m_{i},\Delta_{{\rm ivc}}^\pm\right\} $.

As an example, in Fig.~\ref{fig:particleholesupplemental} we present the full $T=0$ chemical potential phase diagram for a certain choice of interaction parameters. It is this normal-state phase diagram we use for the transport calculations presented in Fig.~1 in the main text.
Due to the particle-hole asymmetry, the DOS in the electron side is much larger as compared to the hole side. This results in stronger flavor-symmetry breaking, larger $\left|\nu\right|=2$ gaps, and additional integer filling gaps on the electron side.

The existence of an insulating state at $\nu=3$, but not at $\nu=-3$ is in line with experimental results. The larger superconducting dome at fillings $\nu=-2-\epsilon$ as compared to  $\nu=2+\epsilon$ which is seen in experiments, can also be understood from Fig.~\ref{fig:particleholesupplemental}. Due to the weaker effective interactions for the holes, the occupation of the two degenerate flavors responsible for the superconductivity extends through a significant portion of the phase diagram and also to higher hole filling.

\begin{figure}
\begin{centering}
\includegraphics[scale=0.6]{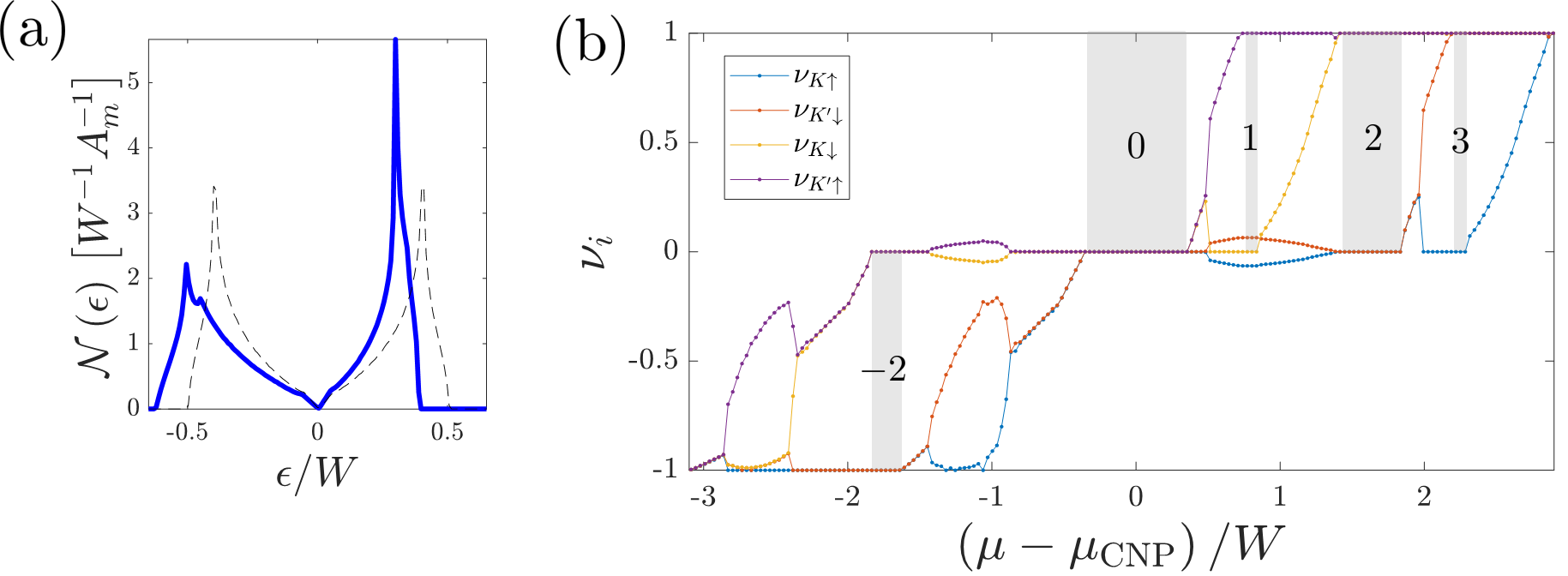}
\par\end{centering}
\caption{\label{fig:particleholesupplemental}
(a)
The DOS used for all calculations in this work (blue solid line). Notice that the hole band is significantly wider as compared to the electron one. The dashed black lines corresponds to the same DOS with $f_{\rm p-h}=0$, i.e., particle-hole symmetric dispersion.
(b)
Mean-field occupation per unit cell $\nu_i$ of individual flavors as a function of chemical potential (relative to the CNP) for $T=0$. The same parameters as in Fig.~1 in the main text were used: $U_{\rm C}=0.72W$, $U_{\delta}=0.2W$, $g_{{\rm ph},1}=g_{{\rm ph},2}=0.12W$. Gray areas mark incompressible regions, and for each one the appropriate integer $\nu$ is denoted.
}
\end{figure}

We also present a full schematic phase diagram of our model at various fillings and temperatures in Fig.~\ref{fig:schematicphasediagram}, which reflects the main features shown Fig.~1 in the main text.
\begin{figure}
\begin{centering}
\includegraphics[scale=0.68]{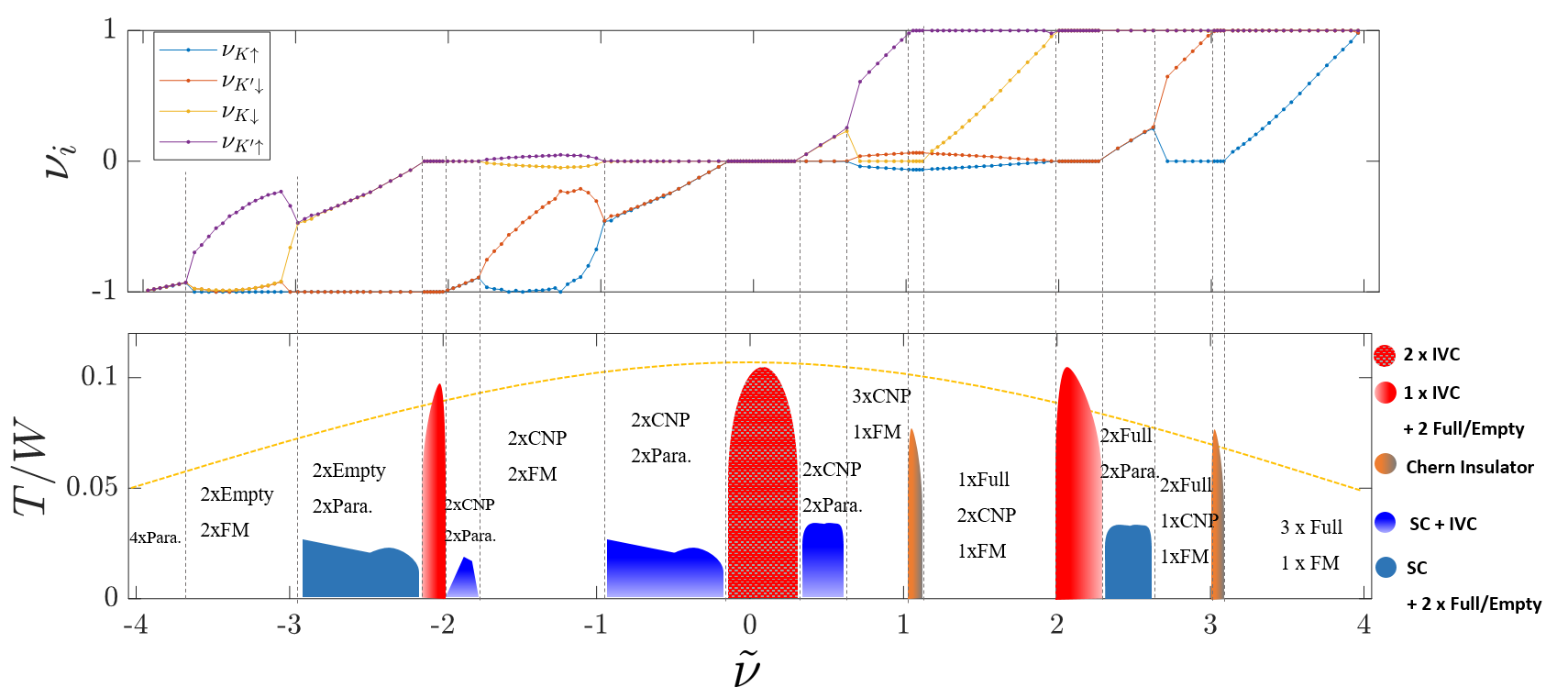}
\par\end{centering}
\caption{\label{fig:schematicphasediagram}
Schematic phase diagram of the model presented. In the upper panel we plot the same mean-field occupation per unit cell data as in Fig.~\ref{fig:particleholesupplemental}, as a function of $\tilde{nu}$. 
The colored shape in the lower panel represent areas of incompressibility or superconductivity, with the legend appearing to the right. In the legend: IVC signals an inter-valley coherent sector, ``Full/Empty'' means a flavor is completely filled/empty, and ``SC'' stand for superconductivity. The vertical dashed gray lines in the figure mark the cascade of flavor symmetry breaking phase transitions, where each phase is labeled accordingly. Here ``CNP'' means a flavor is half-filled (or near the charge-neutrality point), ``Para.'' stands for paramegnetic (i.e., non-polarized states), and ``FM'' stands for a ferromagnetic phase.
The dotted yellow line going over the entire phase diagram marks an approximate crossover temperature, above which the cascade of symmetry breaking transitions is qualitatively different.
}
\end{figure}

\section {High temperature features}

To elucidate some of the features observed in Fig. 1
in the main text, we plot the hole-side mean-field occupation from which Fig.~1 was compiled at two different temperatures, see Fig.~\ref{fig:SuppSCcuts}. At zero temperature, we see a very
similar behavior to Fig.~2b, with the
main difference being at the Fermi level resets (around $\left|\mu-\mu_{{\rm CNP}}\right/W\approx1,2.4$).
Since in the hole side the bandwidth is significantly larger, the Fermi level
does not go down all the way to the Dirac point in this plot. 

At a higher temperature we observe an appreciably different behavior.
First, the low compressibility states at integer fillings (which are completely incompressible at $T=0$) are altogether absent, as one might expect
when reaching high enough temperatures.
Second, the cascade of symmetry breaking is distinctive from the $T=0$ case. 
At low filling, all the flavors begin to fill together, until eventually they split into
two spin-valley locked sector due to the inter-valley antiferromagnetism-inducing
term $g_2$. Then, at intermediate filling, the flavor
symmetry appears to be restored. Eventually, close to full occupation
of the flat bands, we find a miniature cascade of flavor symmetry
breaking, similar in nature to the one discussed in Ref.~\cite{DiracRevivals}.

This distinctive behavior can be understood by observing that (i) the
DOS is gradually increased within each flavor, and (ii) one should
consider the entropy of the itinerant electron (roughly proportional
to the temperature time the DOS at the Fermi level). Namely, the flavor-symmetric
state at intermediate filling gains a significant amount of entropic
free-energy by ``de-polarizing''. This entropic contribution is eclipsed
by the exchange energy at high enough filling (where the DOS is high)
or at low fillings (where the DOS is too diminished to produce a sizable
entropic free-energy). 

The phenomenological model we present thus features phase transitions at intermediate temperatures, where the symmetries of the low and high temperature phases may be rather different.
This is in fact reflected by Fig.~1 in the main text, where different patterns of low compressibility appear above $T\sim0.05W$.
For example, the Fermi level resets shown in Fig.~\ref{fig:SuppSCcuts}b are responsible for the features apparent at the top left corner Fig.~1, as they extend slightly below $T=0.15W$.
However, we note that the specific details of this high temperature phase, e.g., where the compressibility drops are found, are much more sensitive to details of the band structure (whose features we only effectively describe), and should thus be treated with caution.

\begin{figure}
\begin{centering}
\includegraphics[scale=0.6]{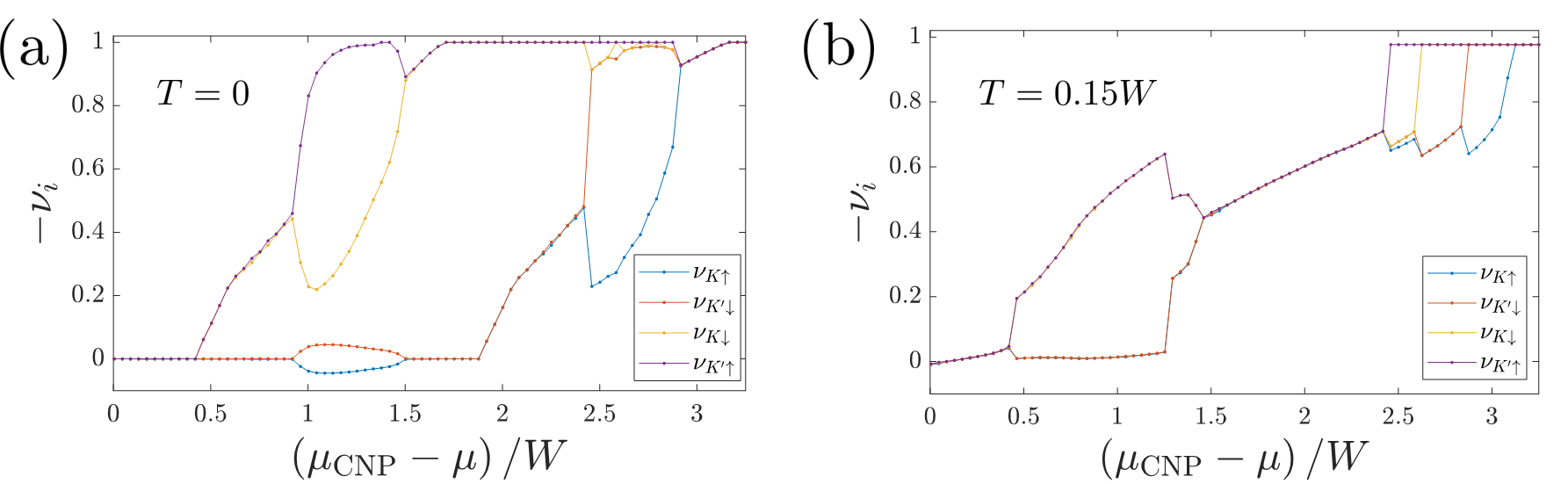}
\par\end{centering}
\caption{\label{fig:SuppSCcuts} Mean-field occupation per unit cell $n_{i}$
of individual flavors in our model as a function of the chemical potential
(relative to the CNP) on the hole-side. Notice the minus sign in both axes, indicating the doping of holes going from left to right. We have used the same parameters of Fig.~1
in the main text: $U_{{\rm C}}=0.7W$, $U_{\delta}=0.15W$, $g_1=g_2=0.12W$.
In panel (a) the temperature is $T=0$, and in panel (b) we used $T=0.15W$.}
\end{figure}

\section{Plotting as a function of $\tilde \nu$} \label{app:nutilde}
The mean-field calculations presented in this work were all performed as a function of chemical potential $\mu$. Instead of plotting measurable quantities as a function of $\mu$, we wish to plot as a function of the gate-voltage $V_G$, which is the experimentally controlled parameter.

Although it is commonly regarded as proportional to the filling itself, $\nu$, it is actually corrected by the quantum capacitance,
\begin{equation}\label{eq:Vgnutilde}
    V_G\left(\nu \right)=\int_{\nu_0}^{\nu}d\nu'\left[C_g^{-1}+\frac{d\mu}{d\nu'} \right],
\end{equation}
where $ V_G\left(\nu_0\right)=0$. The parameter we then plot by is
\begin{equation}
    \tilde \nu = \frac{8}{ V_G\left(\nu=4\right)-V_G\left(\nu=-4\right)}V_G -4\frac{V_G\left(\nu=4\right)+V_G\left(\nu=-4\right)}{V_G\left(\nu=4\right)-V_G\left(\nu=-4\right)},
\end{equation}
closely mimicking the experimental scenario. In Fig.~\ref{fig:supp_nutilde} we show an example of how $\mu$, $\nu$, and $\tilde{\nu}$ relate to one another. Notice that $\nu$ and $\tilde{\nu}$ have a near one-to-one correspondence in metallic regions, yet this gets significantly distorted in incompressible regions, as one might surmise from Eq.~\eqref{eq:Vgnutilde}.

We note that in realistic experiments, effects of disorder may lead to inhomogenous broadening of certain features, e.g., insulating signatures, and further alter the measurements.
\begin{figure}
\begin{centering}
\includegraphics[scale=0.7]{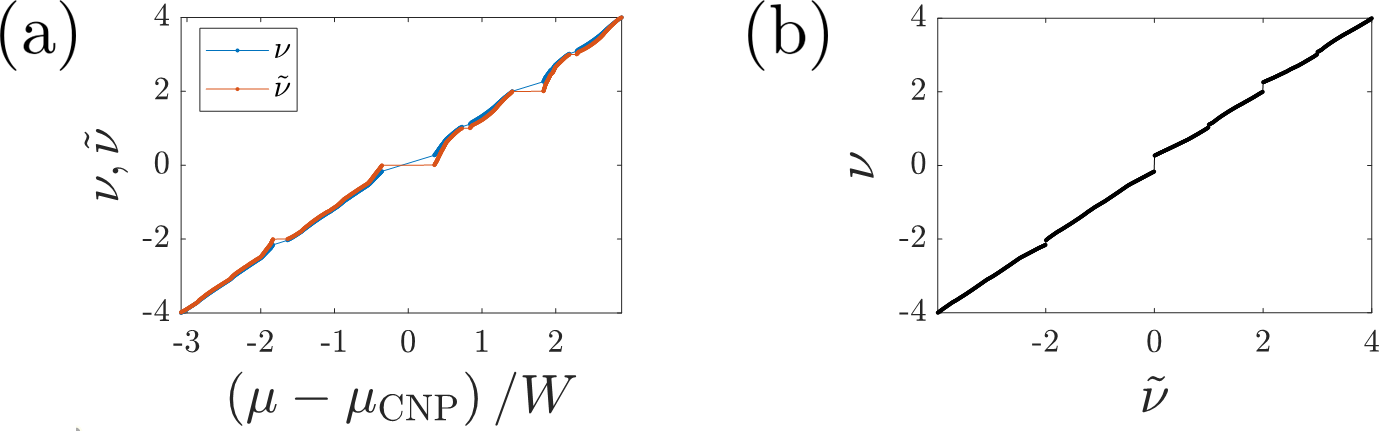}
\par\end{centering}
\caption{\label{fig:supp_nutilde} 
(a) An example for the relation between $\mu$, $\nu$, and $\tilde{\nu}$ for the data presented in Fig.~\ref{fig:particleholesupplemental}b.
(b) The explicit relation between $\nu$ and $\tilde{\nu}$. The correspondence is roughly linear, and gets distorted around the less compressible (or entirely incompressible) regions.
}
\end{figure}

\section {Renormalization Group  equation for superconductivity}\label{app:SCRGequation}
Our starting point for deriving the RG equation Eq.~(6) in the main
text is the partition function
\begin{equation}
Z=\int D\bar{\psi}D\psi e^{-{\cal S}\left[\bar{\psi},\psi\right]},
\end{equation}
with the action
\begin{align}
{\cal S}\left[\bar{\psi},\psi\right] & =\sum_{n,\mathbf{k},\lambda}\left(\xi_{\mathbf{k},\lambda}-i\omega_{n}\right)\bar{\psi}_{n\mathbf{k}\lambda}\psi_{n\mathbf{k}\lambda}\nonumber \\
 & +V\sum_{n,m,\mathbf{k},\mathbf{q}}\bar{\psi}_{m\mathbf{q}1}\bar{\psi}_{\bar m\mathbf{\bar q}2}\psi_{\bar n\mathbf{\bar k}2}\psi_{n\mathbf{k}1},\label{eq:supercondAction}
\end{align}
where $\bar x=-x, x=m,n,\mathbf{q},\mathbf{k}$, the field  $\psi_{n\mathbf{k}\lambda}$ is a fermionic Grassman variable
corresponding to a fermion with Matsubara frequency $\omega_{n}=\left(2n+1\right)\pi T$,
momentum $\mathbf{k}$, and generalized band index $\lambda$. $\xi_{\mathbf{k},\lambda}$
are the mean-field spectra obtained from the variational calculation
within our phenomenological model, and the summation over $\mathbf{k}$
has a cut-off in energy, such that $\left|\xi_{\mathbf{k}}\right|<\Lambda$.
In Eq.~\eqref{eq:supercondAction} we assume that the interaction in the Cooper
channel acts between the two relevant bands with indices $\lambda=1,2,$
e.g., $"1"=K,\uparrow$, and $"2"=K',\downarrow$. We have also kept
only the relevant zero-momentum, zero frequency component of the interaction,
corresponding to a uniform superconducting order parameter.

We perform a Hubbard-Stratonovich transformation to obtain
\begin{equation}
Z=\int D\bar{\Delta}D\Delta D\bar{\psi}D\psi e^{-\tilde{{\cal S}}\left[\bar{\Delta},\Delta,\bar{\psi},\psi\right]},
\end{equation}
\begin{align}
\tilde{{\cal S}}\left[\bar{\Delta},\Delta,\bar{\psi},\psi\right] & =\sum_{n,\mathbf{k},\lambda}\left(\xi_{\mathbf{k},\lambda}-i\omega_{n}\right)\bar{\psi}_{n\mathbf{k}\lambda}\psi_{n\mathbf{k}\lambda}+\frac{1}{V}\bar{\Delta}\Delta\nonumber \\
 & +i\sqrt{\frac{T}{\Omega}}\sum_{n,\mathbf{k}}\left(\bar{\Delta}\psi_{-n,\mathbf{-k},2}\psi_{n\mathbf{k}1}+\Delta\bar{\psi}_{n\mathbf{k}1}\bar{\psi}_{-n,-\mathbf{k},2}\right).
\end{align}
Integrating over the fermions with $\Lambda-d\Lambda<\left|\xi_{\mathbf{k}}\right|<\Lambda$,
we compare the coefficients of the $\bar{\Delta}\Delta$ term in $\tilde{{\cal S}}$,
to find
\begin{equation}
\frac{1}{V\left(\Lambda-d\Lambda\right)}=\frac{1}{V\left(\Lambda\right)}+\frac{T}{\Omega}\sum_{n,\mathbf{k}}^{\Lambda-d\Lambda<\left|\xi_{\mathbf{k}}\right|<\Lambda}\frac{1}{\omega_{n}^{2}+\xi_{\mathbf{k}}^2}.\label{eq:RGfermionicstep}
\end{equation}
where we have assumed $\xi_{\mathbf{k},1}=\xi_{\mathbf{k},2}\equiv\xi_{\mathbf{k}}$.
Assuming the temperature is much smaller than the cutoff, we may replace
the Matsubara sum with an integral to approximate the change in $V\left(\Lambda\right)$ due to the electronic contribution, 
\begin{equation}
\left(dV\right)_{\rm el} \approx{\cal N}\left(\Lambda\right)\frac{d\Lambda}{\Lambda}V^2,\label{eq:RGdosContribution}
\end{equation}
with ${\cal N}\left(\Lambda\right)$ the electronic density of states
at a distance $\Lambda$ away from the Fermi energy.

We have left out of the action in Eq.~\eqref{eq:supercondAction}
the phonon-mediated interaction, $V_{{\rm ph}}\left(\mathbf{q},\omega\right)\propto\frac{1}{\omega^{2}-\omega_{{\rm ph}}^{2}\left(\mathbf{q}\right)}$,
where $\omega_{{\rm ph}}\left(\mathbf{q}\right)$ is the dispersion
of the phonon branch mediating the interaction.
We consider the phonons discussed in Ref.~\cite{LianBioBernevigPRLphonons}, with an acoustic branch which is folded into the mBZ. As a result of this folding, many ``pseudo-optical'' phonon branches are generated, which extend even beyond the flat-band bandwidth $W$. In our analysis, we do not distinguish between these different branches. 

Importantly, the phonon mediated interaction is attractive for $\omega<\omega_{{\rm ph}}$. As we lower
the energy cutoff of our fermionic model, more phonon modes satisfy
this condition and contribute to the attraction. This leads to $V$
becoming increasingly attractive,
\begin{equation}
\left(dV\right)_{\rm ph} =\frac{V^*}{W}d\Lambda,\label{eq:RGphononic}
\end{equation}
where we have implicitly assumed a constant density of states for the phonons, and $\left(-V^{*}\right)$ is the total contribution to the attraction
strength of all the phonons with $0<\omega_{{\rm ph}}<W=\Lambda_{0}$.
We combine the electronic and phononic contributions, Eqs. (\ref{eq:RGdosContribution}), and (\ref{eq:RGphononic}),
to write the flow equation in the form presented in the main text,
\begin{equation}\label{eq:RGsuppequation}
\frac{d}{d\Lambda}V=\frac{{\cal N}\left(\Lambda\right)}{\Lambda}V^{2}+\frac{V^{*}}{W}.
\end{equation}
We note here, as we did in the main text, that Eq.~\eqref{eq:RGsuppequation} suggests a critical $W$ exists for given $V_0$ and $V^*$, below which superconductivity is absent from our calculations. Consider as an extreme example the $W\to 0$ case, where we find that if $V_0-V^*>0$, there can be no superconductivity, since the coupling constant remains repulsive. Too small $W$ thus hinders the retardation mechanism in an intricate way.
Yet, one should keep in mind that a reduction of $W$ should be accompanied also by a modification of $V^*$ and $V_0$, as the phonons above the ``new'' reduced $W$ should be accounted for.

This equation is to be contrasted with that of the frequently used approximation
\begin{equation}
\frac{dV}{d\Lambda}=\frac{{\cal N}\left(\Lambda\right)}{\Lambda}V^{2}+\delta\left(\Lambda-\omega_{D}\right)V_{D},\label{eq:suppRGequationDebye}
\end{equation}
where the attractive retarded interaction $V_{D}$ becomes effective
``all at once'' once the cutoff becomes smaller than the Debye frequency
$\omega_{D}$. 

\begin{figure}
\begin{centering}
\includegraphics[scale=0.65]{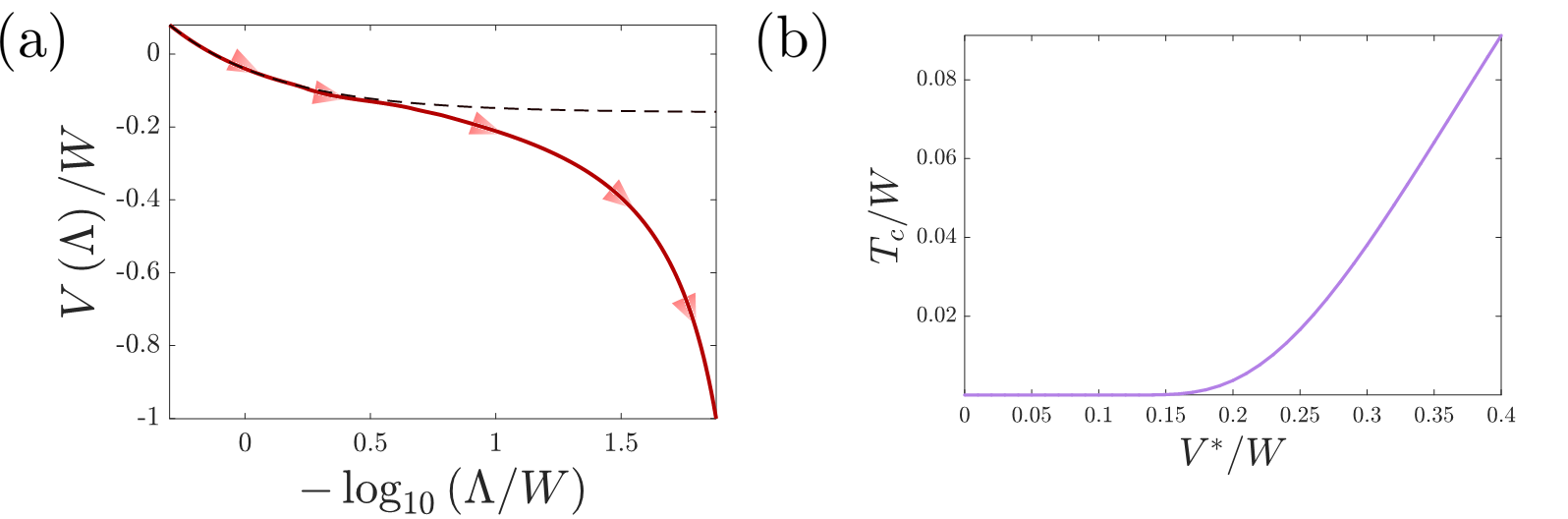}
\par\end{centering}
\caption{\label{fig:suppRGflow} (a) The RG flow of the Cooper channel interaction
$V\left(\Lambda\right)$ (solid red line). The flow without the contribution
of electronic screening, equivalent to taking ${\cal N}\left(\Lambda\right)\left(\Lambda\right)\to0$
in Eq. \eqref{eq:RGsuppequation}, is given by the dashed black line
for comparison. The calculation was performed using the same parameters as in Fig.~1 in the main text, at the approximate filling $\tilde\nu\approx-2.4$. (b) Dependence of $T_{c}$ on the strength of
the phonon-mediated attraction $V^{*}$, showing a roughly linear
behavior in our regime of interest, indicating strong-coupling superconductivity \cite{AllenDynesLinearinLambdaTc}.
The same parameters as in (a) were used, with $V^{*}$ varied.}
\end{figure}

In Fig. \ref{fig:suppRGflow}a we plot the RG flow of $V\left(\Lambda\right)$
at a representative filling. The flow has two distinct regimes. At
the higher cutoff scale, the flow from repulsive to attractive interactions
is mostly due to the phonons gradually contributing. As $\Lambda$
decreases, the electronic contribution takes over, leading to a divergent
coupling constant. This flow highly resembles the well-known Tolmachev-Morel-Anderson
paradigm of superconductivity \cite{tolmachev1962logarithmic,AndersonMorel}.
The dependence of the superconducting $T_{c}$ on the phonon-induced
interaction strength is demonstrated in Fig. \ref{fig:suppRGflow}b,
showing a transition from exponentially small $T_{c}$, consistent
with weak-copling BCS theory, to an approximate linear dependence
on $V^{*}$, indicative of a strong-coupling behavior, as discussed in Ref.~\cite{AllenDynesLinearinLambdaTc}. We stress that
this dependence is a consequence of the effective coupling
constant, i.e., the DOS at the Fermi level times the attraction strength,
being of order unity.

Finally, we address the issue of non-degenerate bands, i.e., $\xi_{\mathbf{k},1}=\xi_{\mathbf{k}}+\delta$, and $\xi_{\mathbf{k},2}=\xi_{\mathbf{k}}-\delta$. If the band splitting is sufficiently small, superconductivity can still be sustained. This happens near some of the superconducting domes in our calculations. For example, the left boundary of the dome near filling $\tilde{\nu}=-2-\epsilon$ is reflects the flavor-polarization shown near $\left(\mu-\mu_{\rm CNP}\right)/W\approx-2.5$.
In such cases, we calculate $T_c$ in the method we have presented for $\delta=0$, and extract the appropriate transition temperature using the implicit relation \cite{maki1969gapless}
\begin{equation}
    \log\frac{T_c\left(\delta\right)}{T_c\left(\delta=0\right)}
    =
    \psi\left(\frac{1}{2}\right)-\psi\left(\frac{1}{2}+\frac{\delta}{2\pi T_c\left(\delta\right)}\right),
\end{equation}
where $\psi\left(x\right)$ is the digamma function.
In this formula, $\delta$ plays the role of a ``pair breaking field'' undermining superconductivity.

\subsection {Analytic solution of the RG equation}
The RG equation we have derived may be analytically solved in three different cases:
(i) When the density of states for the electrons is constant, ${\cal N}={\rm const. }$, (ii) when the electronic density of state is linear in $\Lambda$, and (iii) when the phonon density of states [which is taken constant in Eq.~\eqref{eq:RGsuppequation}] goes like $\Lambda^{-1}$.

For the sake of discussion, we focus on case (ii), and assume
\begin{equation}
{\cal N}\left(\Lambda\right)=\frac{p\Lambda}{W^{2}}.
\end{equation}
This case is of particular importance in the system we study, as the density of states is indeed linear in certain regimes, particularly when the Fermi energy is near the Dirac points. Thus, analytic solutions of Eq.~\eqref{eq:RGsuppequation} may have some qualitative relevance for the study of the MATBG phase diagram.

We find the solution of the differential equation has the form
\begin{equation}
W-\Lambda=W\sqrt{\frac{W}{p V^{*}}}\left[\tan^{-1}\left(\sqrt{\frac{p}{WV^{*}}}V\left(W\right)\right)-\tan^{-1}\left(\sqrt{\frac{p}{WV^{*}}}V\left(\Lambda\right)\right)\right],
\end{equation}
from which we can extract the critical temperature by simplifying
and setting $V\left(\Lambda=T_{c}\right)\to-\infty$, which produces the expression
\begin{equation}
T_{c}/W=1-\sqrt{\frac{W}{p V^{*}}}\left[\tan^{-1}\left(\sqrt{\frac{p}{WV^{*}}}V\left(W\right)\right)+\frac{\pi}{2}\right].\label{eq:TclinearinLambda}
\end{equation}
Eq.~\eqref{eq:TclinearinLambda} provides us some important insights.
First, we can see some trends we expected for $T_{c}$. As $V^{*}$
increases, so does $T_{c}$. This is not surprising, since more phonon-mediated
attraction should naturally lead to more robust superconductivity.
Larger $p$ also boosts $T_{c}$, as the electronic contribution,
which is related to the density of states, is enhanced. Conversely,
a larger initial repulsion $V\left(W\right)$ suppresses $T_{c}$.

More importantly, the right hand side of Eq.~\eqref{eq:TclinearinLambda}
may become negative, indicating the absence of a superconducting instability.
More concretely, this points to the existence of a critical phonon-mediated
attraction $V_{c}^{*}$, below which superconductivity vanishes. Taking
the bare repulsion $V\left(W\right)\to0$, we can get an estimate
on this critical interaction strength,
\begin{equation}
V_{c}^{*}/W\approx\frac{\pi^{2}}{4p}.
\end{equation}
The sizable critical interaction, which is of the order of $W$, is
due to the vanishing density of states at $\Lambda\to0$.
We point out that this result is
reminiscent of Ref.~\cite{GrapheneBCS}, where it was found that
the critical dimensionless coupling constant for BCS superconductivity
[akin to solving Eq.~\eqref{eq:suppRGequationDebye}] in graphene
at the charge neutrality point is unity. 
This consequence is consistent
with our treatment of the presented phenomenological model, which features no superconductivity
near a Dirac point. The superconductivity
in this work in the vicinity of integer fillings owes its existence to the symmetry-broken state, where
electrons have a finite DOS at the Fermi level.  

\section {Superconducting phase fluctuations}\label{appsec:SCfluctuations}
We follow Ref.~\cite{HalperinSuperconductorResistance} to elucidate the role of phase fluctuations, which lead to the Berezinskii-Kosterlitz-Thouless (BKT) phase transition, on transport. 
The critical temperature we derive from the RG flow equation will be labeled here as $T_{c}$, coinciding with the ``bare'' Ginzburg-Landau transition temperature, i.e., the critical temperature once phase fluctuations are neglected.

On the other hand, we have the BKT transition temperature $T_{\rm BKT}$, which is related to $T_c$ by
\begin{equation}
    T_{\rm BKT}=\frac{T_c}{1+\tau_c},
\end{equation}
where  $\tau_c$ is a dimensionless parameter ordinarily much smaller than unity, parameterizing the role of phase fluctuations of the superconducting order parameter.
Its value may be evaluated from microscopic parameters in the clean- and dirty-superconductor limits (where in the clean limit the mean-free-path $\ell$ is much larger than the Ginzburg-Landau correlation length evaluated at $T_c$,  $\xi_c$, and vice-versa) \cite{abrikosov2017fundamentals,HalperinSuperconductorResistance}, 
\begin{equation}
    \tau_{c}=\begin{cases}
   {T_{c}}/{T_{F}} & {\rm clean}\\
    {0.14}/{k_{F}\ell} & {\rm "dirty"}
    \end{cases},
\end{equation}
with $T_F$ the Fermi temperature, and $k_F$ the Fermi momentum.

Examination of the experimental normal-state resistance \cite{CaoUnconventionalSC,YankowitzTuningMATBG}, as well as the superconducting correlation length \cite{CaoUnconventionalSC}, leads to the estimate $\ell \sim 5 \xi_c$, i.e., the experimental situation appears to be closer to the clean limit. Hence, the quantity ${T_{c}}/{T_{F}}$ will essentially determine the role and importance of phase fluctuations in transport.

The fluctuation correlation length, denoted by $\xi^{*}$, may be
evaluated from the following formula for $T>T_{{\rm BKT}}$ \cite{HalperinSuperconductorResistance},
\begin{equation}
\xi^{*}=b^{-1/2}\xi_{c}\sinh\sqrt{b\frac{T_{c}-T_{{\rm BKT}}}{T-T_{{\rm BKT}}}},\label{eq:suppCorrelationlength}
\end{equation}
where $b$ is a dimensionless constant of order unity. Notice Eq.~\eqref{eq:suppCorrelationlength} is also valid below $T_{c}$. 
In fact, it is apparent that the phase-fluctuation dominant regime is the one
where $T_{{\rm BKT}}<T\lesssim T_{c}$. In other words, the parameter
$\tau_{c}$ determines the width of a temperature ``window'' where
fluctuations become important.

The contribution of superconducting fluctuations to the conductivity
above $T_{{\rm BKT}}$ may be evaluted as \citep{ASLAMASOV1968238}
\begin{equation}
\delta\sigma_{s}=\frac{e^{2}}{h}\frac{\pi}{8\tau_{c}}\left(\frac{\xi^{*}}{\xi_{c}}\right)^{2}.
\end{equation}
We note that in the dirty-superconductor limit, one may write this
correction in the form $\delta\sigma_{s}\approx0.37\sigma_{n}\left(\xi^{*}/\xi_{c}\right)^{2}$,
with $\sigma_{n}$ the normal-state conductivity. We then arrive at
an expression for the resistance,
\begin{equation}
R=\sigma_{{\rm tot}}^{-1}=\frac{h}{e^{2}}\left[hD\frac{\partial n_{{\rm tot}}}{\partial\mu}+\frac{\pi}{8\tau_{c}}\left(\frac{\xi^{*}}{\xi_{c}}\right)^{2}\right]^{-1},
\end{equation}
with $D$ the diffusion constant controlling the normal state resistivity.
In plotting Fig.~1 in the main text we used $hD=1 \times A_m \times W$.

Finally, let us evaluate $T_{F}$ within our mean-field normal-state
solution. For simplicity, let us assume a (possibly gapped) Dirac-like
dispersion, such that
\begin{equation}
E_{\mathbf{k}}=\pm W\sqrt{\left(\frac{\left|\mathbf{k}\right|}{Q}\right)^{2}+\left(\frac{\Delta_{{\rm ivc}}}{W}\right)^{2}}-\mu_{{\rm MF}},
\end{equation}
where $W$ is the bandwidth, $Q\sim A_{m}^{-1/2}$ is the momentum
cutoff, $\Delta_{{\rm ivc}}$ is the normal-state IVC gap, and $\mu_{{\rm MF}}$
is the mean-field chemical potential. Defining the effective mass
$m^{*}=k_{F}/v_{F}$, with $v_{F}=\left|\partial_{\mathbf{k}}E_{\mathbf{k}}\right|_{\left|\mathbf{k}\right|=k_{F}}$,
we find the Fermi temperature $T_{F}=\left(\pi\hbar^{2}n^{*}\right)/\left(m^{*}k_{B}\right)$
(with $n^{*}$ the carrier density),
\begin{equation}
T_{F}=\left|\mu_{{\rm MF}}\right|\left[1-\left(\frac{\Delta_{{\rm ivc}}}{\mu_{{\rm MF}}}\right)^{2}\right].
\end{equation}\label{eq:FermitemperatureTF}

\section {Phase diagram with smaller Coulomb repulsion}\label{appsec:phasediagsmallerU}
As we mention in the main text, a significant amount of experimental work has been carried out where the electrostatic screening in MATBG was manipulated~\cite{YankowitzTuningMATBG,EfetovTuningSC,YoungTuningSC,BLGscreening}.
It is interesting to study how the phenomenological model presented behaves under suppression of the Coulomb repulsion.
This can be done by decreasing both $U_{\rm C}$, which is the dominant interaction meant to represent the structure-less part of the Coulomb repulsion, and $U_{\delta}$, whose strength should also be proportional to the Coulomb interaction (yet its structure reflects the existence of some form-factors in the interaction term).

We show an example of how the phase diagram changes in Fig.~\ref{fig:lessrepulsion}, where for convenience we also show the main result of Fig. 1 from the main text.
One notices several important differences.
First, the correlated insulators at even fillings have all ``weakened'', i.e., became narrower and with lower critical temperatures.
Additionally, the insulators at positive integer fillings have vanished, leaving behind regions of low compressibility as the active bands' Fermi energy is close to the Dirac point.
This effect is expected: the correlated insulators at integer fillings are mostly driven by $U_{\rm C}$, and reducing it makes the insulating phases less favorable in energy as compared to the compressible ones.

The second effect is that the superconducting domes have all widened, and slightly increased their respective $T_c$.
Since the superconducting domes are all cut-off on one of their sides by a spontaneous flavor-symmetry breaking transition, their widening upon reduction of the repulsion is also well-understood. Higher-symmetry phases persist longer in the phase diagram when the repulsion is weaker, leading to larger regions of superconductivity.

An increase in $T_c$ is also expected, as the effective initial coupling constant $V_0$ which goes into the calculation of $T_c$ is more positive when $U_{\rm C}$ is larger.
The fact that the increase in critical temperature is only modest is due to the decrease in $U_\delta$, which has the opposite effect on $V_0$.
We note that a slight increase in superconducting $T_c$ as a result of suppressed Coulomb repulsion is consistent with the results of Ref.~\cite{BLGscreening}.

The effects shown in Fig.~\ref{fig:lessrepulsion} are in agreement with Refs.~\cite{YankowitzTuningMATBG,EfetovTuningSC,YoungTuningSC}, which featured a common trend: more screening, i.e., weaker repulsion, leads to less insulators and more superconducting regions in the phase diagram.

\begin{figure}
\begin{centering}
\includegraphics[scale=0.75]{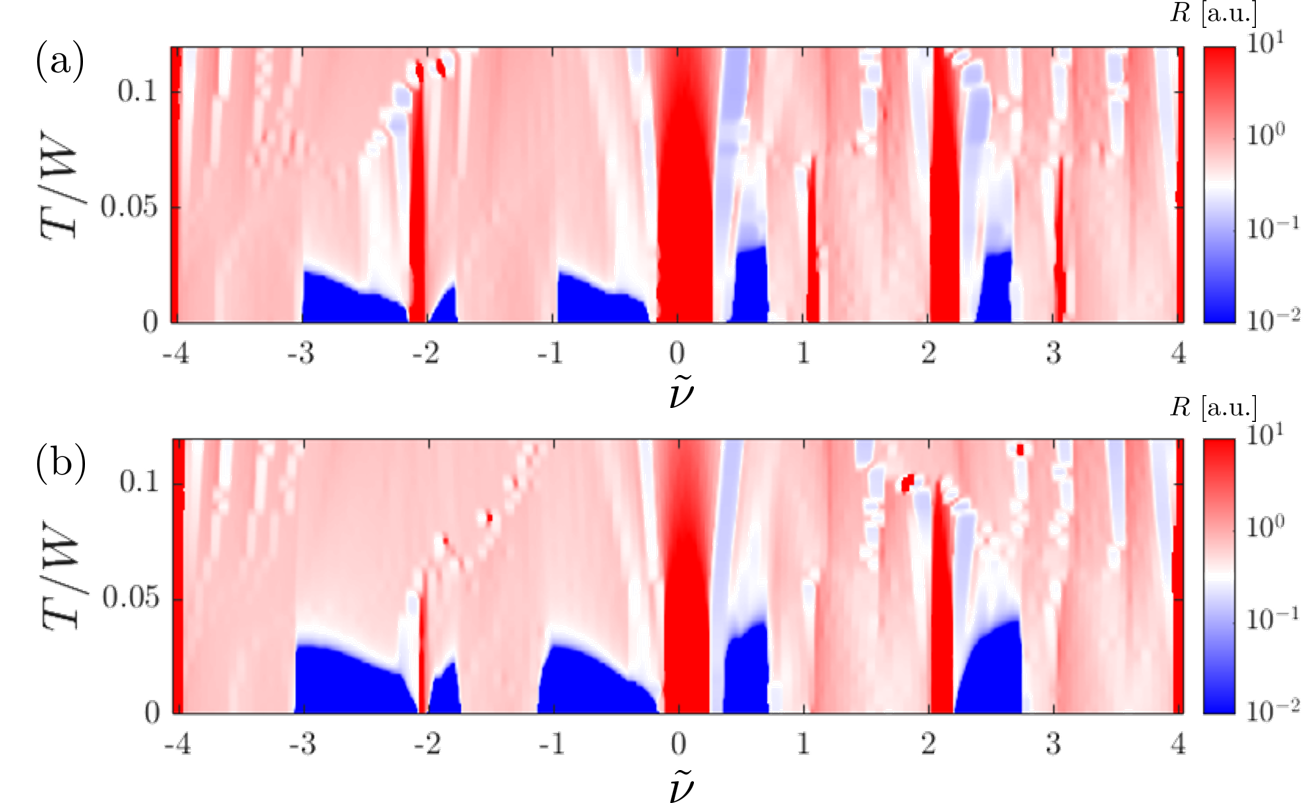}
\par\end{centering}
\caption{\label{fig:lessrepulsion}
The effect of slightly smaller effective Coulomb interactions on the phase diagram.
(a) 
The phase diagram presented in Fig. 1 in the main text. The parameters used here were: $U_{{\rm C}}=0.7W$, $U_{\delta}=0.15W$, $g_1=g_2=0.12W$, and $V^{*}=0.24W$.
(b)
The same phase diagram calculated with the same parameter values, except for $U_{\rm C}=0.62W$ and $U_\delta=0.13W$.
}
\end{figure}

\section {Characteristic values of coupling constants}\label{appsec:couplingvalues}
After establishing the phenomenological model presented and analyzed, and obtaining a phase diagram which closely resembles experimental results, one may ask whether the \textit{values} of the coupling constants used to obtain this diagram are consistent with theoretical predictions and experimental measurements.

Let us begin with addressing the bandwidth of the flat bands, $W$.
Theoretical predictions based on the continuum Bistritzer-Macdonald model~\cite{BistritzermacdonaldPNAS} and accounting for lattice relaxation effects, estimate the non-interacting bandwidth near the magic angle may be as small as $3-5$ meV.
However, it has been argued~\cite{BandStructurenoramlizations,HartreeBandStructure,Mcdonaldweakfieldhall,FillingDependentRenormalization} that Coulomb interactions strongly renormalize the bandwidth increasing it to an order of
\begin{equation}
    W \sim 20-30 {\rm meV}.
\end{equation}
This estimate appears to be more consistent with compressibility and tunneling measurements performed near the magic angle~\cite{DiracRevivals,YazdaniRevivals}.

Next, we consider the most dominant interaction in our model, $U_{\rm C}$.
A rough estimate for its strength may be obtained by $U_{\rm C}\approx{e^2/r_{\rm eff}}$, with $r_{\rm eff}$ being the length scale determining the size of the interactions. In the absence of screening from nearby gates (or if the gates are at a distance greater than $r_{\rm eff}$), we may approximate $r_{\rm eff}\approx \epsilon a_c/\sin \theta$, where $\epsilon$ is the dielectric constant of the h-BN substrate which is of the order $\epsilon\sim {\cal O}\left(10\right)$, $a_c=0.245$ nm is the graphene lattice constant, and $\theta$ is the twist angle ($a_c/\sin\theta$ is the moir\'e lattice constant).
These lead to a characteristic interaction energy scale of
\begin{equation}
    U_{\rm C} \sim 15-25 {\rm meV}.
\end{equation}
We mention that the experimental results in Ref.~\cite{DiracRevivals} were remarkably reproduced by the authors of that work using a simplified model with only symmetric $U_{\rm C}$-like interactions on the same order of $W$, just as our estimates thus far suggest.

Moving on to the form-factor-related contribution to the interaction Hamiltonian, $U_\delta$, we rely on results obtained by Ref.~\cite{BultnickKhalaf2020}, which found
\begin{equation}
    U_\delta \sim 4-6 {\rm meV}.
\end{equation}

The effectively-instantaneous interactions mediated by optical phonons can be estimated based on Ref.~\cite{MacdonaldPRLopticalPhonons}, 
\begin{equation}
   g_1 \sim g_2 \sim 1-2 {\rm meV}.
\end{equation}

Finally, we may evaluate the contribution to the retarded electron-electron attraction from phonons at energy scales up to $W$, as described by the coupling constant $V^*$.
Ref.~\cite{LianBioBernevigPRLphonons} estimates the contribution of phonons up to a Debye-like energy scale of $\omega_D\approx2$ meV to be of order $\sim 1$ meV.
Using the very crude estimate of a constant density of states of phonons contributing to the interaction, we may estimate $V^* \sim 1 {\rm mev}\frac{W}{\omega_D}$.
Treating this approximation conservatively, since phonons at high energy scales are expected to play a smaller part in superconductivity, we estimate the order of magnitude
\begin{equation}
    V^* \sim 5-10 {\rm mev}.
\end{equation}

To conclude, we bring here for the sake of convenience the parameter values used in calculating Fig. 1 in the main text, which are all consistent with our estimates above:
$U_{{\rm C}}=0.7W$, $U_{\delta}=0.15W$, $g_1=g_2=0.12W$, and $V^{*}=0.24W$.
Obtaining an experimentally-consistent phase diagram using these well-justified values of the coupling constants adds further credibility to the model presented in this work. 

\section {Suppression of superconductivity by h-BN alignment}\label{appsec:hbn}
Let us consider a scenario in which explicit sublattice-symmetry breaking
is introduced via changing the single-particle part of the Hamiltonian,
\begin{equation}
h_{0}\left(\mathbf{k}\right)\to h_{AB}\left(\mathbf{k}\right)=h_{0}\left(\mathbf{k}\right)+\Delta_{AB}\sigma_{z},
\end{equation}
due to alignment with one of the encapsulating h-BN subsrtrates. $h_{AB}$
is diagonalized by the transformation
\begin{equation}
h_{D}={\cal U}^{\dagger}h_{AB}{\cal U},
\end{equation}
with $h_{D}$ a diagonal matrix, and

\[
{\cal U}=\begin{pmatrix}\cos\theta & -\sin\theta e^{i\phi\tau_{z}}\\
\sin\theta e^{-i\phi\tau_{z}} & \cos\theta
\end{pmatrix},
\]
where ${\cal U}$ is written in the sublattice basis, and

\[
\cos2\theta=\frac{\Delta_{AB}}{\sqrt{\left|\epsilon_{\mathbf{k}}\right|^{2}+\Delta_{AB}^{2}}},\,\,\,\sin2\theta=\frac{\left|\epsilon_{\mathbf{k}}\right|}{\sqrt{\left|\epsilon_{\mathbf{k}}\right|^{2}+\Delta_{AB}^{2}}},\,\,\,\tan\phi=\frac{f_{y}\left(\mathbf{k}\right)}{f_{x}\left(\mathbf{k}\right)}.
\]

It is now convenient to work in the conduction/valence band basis
instead of the sublattice-polarized basis, i.e., working with $\Phi={\cal U}^{\dagger}\Psi$,
and projecting interactions onto one of the bands, e.g., conduction
band. For the purpose of illustration, it is sufficient to consider
the Cooper channel interaction term 
\begin{align*}
U_{\sigma_{x}} & \equiv u\sum_{\mathbf{k,q}}\Psi^{\dagger}\left(\mathbf{k}\right)\sigma_{x}\Psi\left(\mathbf{q}\right)\Psi^{\dagger}\left(\mathbf{-k}\right)\sigma_{x}\Psi\left(\mathbf{-q}\right)\\
 & =u\sum_{\mathbf{k,q}}\Phi^{\dagger}\left(\mathbf{k}\right){\cal U}_{\mathbf{k}}^{\dagger}\sigma_{x}{\cal U}_{\mathbf{q}}\Phi\left(\mathbf{q}\right)\Phi^{\dagger}\left(\mathbf{-k}\right){\cal U}_{\mathbf{-k}}^{\dagger}\sigma_{x}{\cal U}_{\mathbf{-q}}\Phi\left(\mathbf{-q}\right),
\end{align*}
where upon projecting to the conduction band (with fermionic annihilation
operators $\phi_{c}$), we find the interaction
\begin{equation}
\sum_{\mathbf{k,q}}\tilde{u}\phi_{c}^{\dagger}\left(\mathbf{k}\right)\phi_{c}\left(\mathbf{q}\right)\phi_{c}^{\dagger}\left(\mathbf{-k}\right)\phi_{c}\left(\mathbf{-q}\right),
\end{equation}
and 
\begin{equation}
\tilde{u}=u\left(1-\cos2\theta_{\mathbf{k}}\cos2\theta_{\mathbf{q}}\right).
\end{equation}
 For simplicity, we limit our discussion to the close vicinity of
the Fermi surface, approximating $\epsilon_{\mathbf{k}}\approx\epsilon_{\mathbf{q}}\approx\sqrt{E_{F}^{2}-\Delta_{AB}^{2}}$,
leading to the approximate attenuation of sublattice-scattering interactions
by
\begin{equation}
\tilde{u}/u=1-\left(\frac{\Delta_{AB}}{E_{F}}\right)^{2}.\label{eq:suppAttFactor}
\end{equation}

Let us briefly summarize our findings. In the Cooper channel, the $U_{{\rm \delta}}$ and $g_{{\rm ph},1}$ interactions scatter a pair with a certain sublattice label to an opposite sublattice pair, thanks to the presence of $\sigma_{x,y}$ in the interaction. This enhances superconductivity.
Alignment of the h-BN substrate introduces a $\sigma_z$ term to the Hamiltonian, thereby suppressing these pairing-friendly scattering events, as the $A-B$ balance is broken. Thus, the initial electron-electron repulsion is effectively stronger, and superconductivity is suppressed.

In Fig.~\ref{fig:Supphbn} we show the impact of the attenuation factor
in Eq.~\eqref{eq:suppAttFactor} on the superconducting $T_{c}$. We
start with an initial repulsive interaction in the Copper channel,
\begin{equation}
\tilde{V_{0}}=\frac{U_{{\rm C}}}{2}-\tilde{u}/u\left(U_{{\rm \delta}}+\left|g_{{\rm ph},1}\right|\right),
\end{equation}
and monitor $T_{c}$ at a certain filling. We note that in order to
maintain the same filling with different $\Delta_{AB}$ we modify
also the Fermi level. As we show in Fig.~\ref{fig:Supphbn}, there
is a gradual decrease in the critical temperature with increasing
$\Delta_{AB}$, until it vanishes at some critical value of the sublattice
symmetry breaking potential. The absence of superconductivity in experiments
done on MATBG aligned with h-BN may then be explained by this mechanism,
as the gap opened by h-BN alignment may reach an order of 15-30 meV
\citep{hBNgapsize}, which is comparable with the bandwidth of the
MATBG flat bands. 

\begin{figure}
\begin{centering}
\includegraphics[scale=0.6]{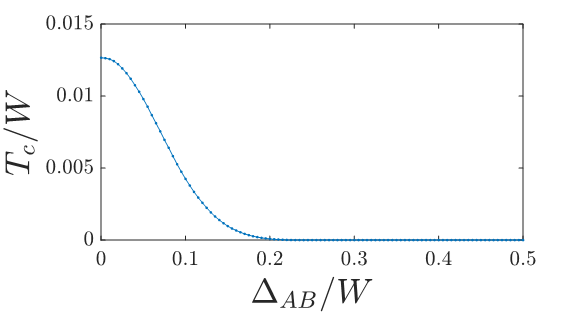}
\par\end{centering}
\caption{\label{fig:Supphbn} Superconducting $T_{c}$ as a function of $\Delta_{AB}$,
calculated at a constant filling of $n_{{\rm tot}}=2.4$, i.e., two
completely filled flavors, and two degenerate opposite-valley flavors,
each $1/5$ filled. Parameters used : $U_{{\rm C}}=0.6W$, $U_{{\rm \delta}}=\left|g_{{\rm ph},1}\right|=0.1W$,
and $V^{*}=0.25W$. }
\end{figure}
\end{widetext}

\end{document}